\documentclass{aa}

\input psfig.sty
\usepackage{psfig}

\font\twlbfs=cmbxsl10 scaled \magstep3

\def\aap{{A\&A}}
\def\aapl{{A\&A} (Letters)}
\def\aaps{{A\&AS}}

\def\aj{{AJ}}

\def\apj{{ApJ}}
\def\apjl{{ApJ} (Letters)}
\def\apjs{{ApJS}}

\def\ga{\mathrel{\hbox{\rlap{\hbox{\lower4pt\hbox{$\sim$}}}\hbox{$>$}}}}

\def\la{\mathrel{\hbox{\rlap{\hbox{\lower4pt\hbox{$\sim$}}}\hbox{$<$}}}}

\def\mnras{{MNRAS}}
\def\msai{{Mem. Soc. Astron. Italia}}
\def\na{{New Astronomy}}
\def\nar{{New Astronomy Reviews}}

\def\pasp{{PASP}}

\def\ssr{{Space Sci. Rev.}}

\newcommand{\xray}{\mbox{X-ray\ }}
\newcommand{\xrays}{\mbox{X-rays\ }}

\newcommand{\be}{\begin{equation}}
\newcommand{\ee}{\end{equation}}
\newcommand{\bea}{\begin{eqnarray}}
\newcommand{\eea}{\end{eqnarray}}

\newcommand{\ASCA}{{\sl ASCA} }

\newcommand{\ROSAT}{{\sl ROSAT} }

\newcommand{\XMM}{{\sl XMM-Newton} }

\newcommand{\gired}{$\chi^2_{\rm red}$}

\newcommand{\Msolar}{\mbox{\,$\rm M_{\odot}$}}        % solar mass
\newcommand{\Mdot}{\stackrel{.}{M}}

\newcommand {\Hei}{He\,{\sc i} }

\newcommand {\Civ}{C\,{\sc iv} }
\newcommand {\Cv}{C\,{\sc v} }
\newcommand {\Cvi}{C\,{\sc vi} }

\newcommand {\Ovii}{O\,{\sc vii} }
\newcommand {\Oviii}{O\,{\sc viii} }

\newcommand {\Sixiii}{Si\,{\sc xiii} }
\newcommand {\Sixiv}{Si\,{\sc xiv} }

\newcommand {\Neix}{Ne\,{\sc ix} }
\newcommand {\Nex}{Ne\,{\sc x} }

\newcommand {\Mgxi}{Mg\,{\sc xi} }
\newcommand {\Mgxii}{Mg\,{\sc xii} }

\newcommand {\Sxvi}{S\,{\sc xvi} }

\newcommand {\Sxv}{S\,{\sc xv} }

\newcommand {\Fexvii}{Fe\,{\sc xvii} }
\newcommand {\Fexviii}{Fe\,{\sc xviii} }

\newcommand {\Fexxv}{Fe\,{\sc xxv} }

\newcommand {\Fek}{Fe\,{\sc -k} }

\newcommand {\Arxvii}{Ar\,{\sc xvii} }
\newcommand {\Arxviii}{Ar\,{\sc xviii} }
\newcommand {\Caxx}{Ca\,{\sc xx} }
\newcommand {\Caxix}{Ca\,{\sc xix} }

\begin{document}

\title
 {Wind clumping and the wind-wind collision zone in the Wolf-Rayet
binary $\gamma^2$\,Velorum
 \thanks{Based on observations obtained with
 {\sl XMM-Newton}, an ESA science mission with instruments and
 contributions directly funded by ESA Member States and the USA (NASA).}
 }
\subtitle{{\twlbfs XMM-Newton} observations at high and low state}
\author{H.~Schild\inst{1}
  \and M.~G\"udel\inst{2}
  \and R.~Mewe\inst{3}
  \and W.~Schmutz\inst{4}
  \and A.J.J. Raassen\inst{3,5}
  \and M.~Audard\inst{6}
  \and T.~Dumm\inst{1}
  \and K.A.~van der Hucht\inst{3,5}
  \and M.A.~Leutenegger\inst{6}
  \and S.L.~Skinner\inst{7}
}

\institute{Institut f\"ur Astronomie, ETH-Zentrum, CH 8092 Z\"urich, Switzerland \\
           e-mail: {\tt hschild@astro.phys.ethz.ch}
 \and      Paul Scherrer Institut, W\"urenlingen \& Villigen,
           CH 5232 Villigen PSI, Switzerland                          \\
           e-mail: {\tt guedel@astro.phys.ethz.ch}
 \and      SRON National Institute for Space Research,
           Sorbonnelaan 2, NL-3584\,CA Utrecht, the Netherlands       \\
           e-mail: {\tt a.j.j.raassen@sron.nl; k.a.van.der.hucht@sron.nl; r.mewe@sron.nl}
 \and      Physikalisch-Meteorologisches Observatorium Davos,
           Dorfstrasse 33, CH 7260 Davos Dorf, Switzerland            \\
           e-mail: {\tt w.schmutz@pmodwrc.ch}
 \and      Astronomical Institute "Anton Pannekoek",
           Kruislaan 403, NL-1098 SJ Amsterdam, the Netherlands       \\
           e-mail: {\tt raassen@science.uva.nl}
 \and      Columbia Astrophysics Laboratory,
           Columbia University, 550 West 120th Street, New York, NY\,10027, USA \\
           e-mail: {\tt audard@astro.columbia.edu; maurice@astro.columbia.edu}
 \and      Center for Astrophysics and Space Astronomy,
           University of Colorado, Campus Box 389, Boulder, CO 80309-0389, USA \\
           e-mail: {\tt skinner@origins.colorado.edu}
          }

\offprints{M.\,G\"udel}
\date{version {\today}. ~~~Received / Accepted}

\abstract{
We present \XMM observations of $\gamma^2$ Velorum (WR\,11,
WC8+O7.5III, $P$\,=\,78.53\,d),
a nearby Wolf-Ray binary system, at its \xray high and low states.
At high state, emission from a hot collisional
plasma dominates from about 1 to 8 keV. At low state, photons between
1 and 4\,keV are absorbed. The hot plasma
is identified with the shock zone between the winds of the primary
Wolf-Rayet star and the secondary O giant. The absorption at
low state is interpreted as photoelectric absorption in the Wolf-Rayet
wind. This absorption  allows us to measure the absorbing column density
and to derive a mass loss rate
$\Mdot$\,=\,8$\times$10$^{-6}$\,M$_\odot$yr$^{-1}$ 
for the WC8 star. 
This mass loss rate, in conjunction with a previous Wolf-Rayet wind model,
provides evidence for a clumped WR wind. A clumping factor of 16 is
required.
The \xray spectra below 1 keV (12~\AA)
show no absorption and are essentially similar in both states.
There is a rather clear separation
 in that emission from a plasma hotter than 5\,MK is heavily absorbed in low state
while the cooler plasma is not.
 This cool plasma must come from a much more extended region
than the hot material. The Neon 
abundance in the \xray emitting material is 2.5 times
the solar value.
The unexpected detection of C~\textsc{v} (25.3~\AA) and C~\textsc{vi} (31.6~\AA)
radiative recombination continua at both phases indicates the presence of a cool
($\sim$\,40,000~K) recombination region located far out in the
binary system. 

}

\titlerunning{\XMM observations of $\gamma^2$\,Velorum (WC8+O7.5III)}
\authorrunning{H. Schild et al.}
\maketitle
%\markboth{}{}

\section{Introduction\label{sec_intro}}

The massive Wolf-Rayet binary system $\gamma^2$~Velorum (WR\,11,
WC8+O7.5III, $P$\,=\,78.53\,d) is an astrophysical
 laboratory in which many
aspects of mass loss and wind-wind collision
phenomena can be studied. The system is relatively nearby,
its {\sl Hipparcos} distance
is $d\,=\,258\pm35$\,pc (van der Hucht et al. 1997; Schaerer et al. 1997).
Both stars in the binary have been recently investigated with
sophisticated model atmospheres and their stellar parameters are reasonably
well known (De Marco \& Schmutz 1999, De Marco et al. 2000).

The binary orbit has been re-determined by Schmutz et
al. (1997) who combined recent with earlier observations 
(Niemela \& Sahade 1980; Pike et al.
1983; Moffat et al. 1986; Stickland \& Lloyd 1990). The orbit is
mildly eccentric and has an
inclination of 63$^\circ$\,$\pm$\,8$^\circ$ (De Marco \& Schmutz 1999).
Because of the high orbital inclination, any
emitting structures are seen through changing absorption columns as
the stars revolve.

Since $\gamma^2$~Vel is the nearest WR star, (see
van der Hucht 2001), it is relatively bright and well
observable at any wavelength, in particular in the X-ray domain.  It has
been observed by all previous X-ray observatories, from the {\sl Einstein}
observatory (White \& Long 1986; Pollock 1987) to {\sl ASCA} (Stevens et
al. 1996; Rauw et al. 2000), and, more recently, by {\sl Chandra}
(Skinner et al. 2001).  Its X-ray observational history has been reviewed
by van der Hucht (2002) and Corcoran (2003).

%%%%% FIGURE 1 %%%%%
\begin{figure}
\hbox{\hskip-0.2cm\psfig{figure=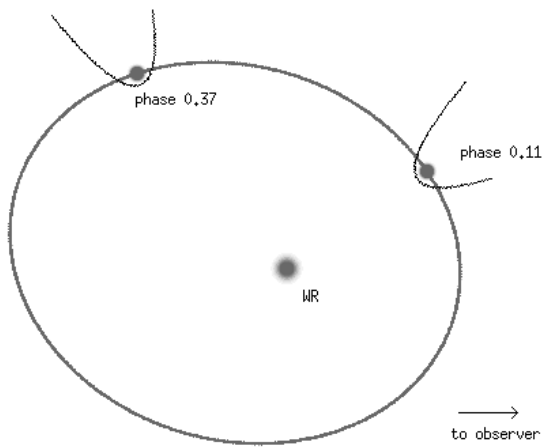,height=7.7cm,width=8.5cm,clip=}}
\caption{Sketch of the $\gamma^2$\,Vel orbital configuration at the
phases of our X-ray observations. The Wolf-Rayet star is at the center.
Shape and orientation of the wind blown cavities around the O star
are schematic only.}
\label{figure1}
\end{figure}

With a series of \ROSAT observations covering the binary orbit of
$\gamma^2$\,Vel, Willis et al. (1995) discovered that the X-ray emission
is a factor of $\sim$\,4 enhanced during a brief time span when the
O-type component is in front.  They also showed that
the steep increase takes place only in \ROSAT's hard \xrays.
They convincingly
interpreted the variable \xray emission to arise from colliding
stellar winds.  The enhancement is explained by the viewing geometry, when
the collision zone can be seen through a rarefied cavity that builds around
and behind the O-type component (see Fig.\,\ref{figure1}). At other phases the
dense WR wind absorbs the
X-rays from the collision zone. The wind blown
cavity is generally orientated 
away from the WC component but it is also somewhat warped because
of the binary motion of the O star.

Here we present \XMM observations of $\gamma^2$\,Vel, taken at two phases.
The first phase is at the maximum X-ray flux, a few days after the
O-type component 
passed 
in front of the WR star.  The second phase is intermediate
between quadrature and superior
conjunction. In this configuration the O star is seen through a 
large
portion of the extended WR atmosphere (Fig.~\ref{figure1}).

 After describing the observations and the
most interesting spectral features we analyze the data in two different ways.
First, we simply take the \xray emitting zone as a source of
light with which the WR wind is irradiated.  The observed
absorption changes between different orbital phases provide unique
information about
the structure of the WR wind.  Secondly, we interpret the X-ray emission
at both phases by a spectral fitting procedure. This reveals new insights into
the geometric and thermal structure and the elemental composition
of the wind-wind collision zone.

\section{Observations\label{Sec_obs}}

The log of our \XMM observations of $\gamma^2$\,Vel is presented in
Tab.~\ref{obslog}.
A description and a preliminary analysis of
the observations are given by Dumm et al. (2003).
Technical information on \XMM and its X-ray instrumentation can be found
in den Herder et al. (2001), Jansen et al. (2001), Str\"uder et al.
(2001), and Turner et al. (2001).

%%%%% TABLE 1 %%%%%
\begin{table}[h]
\caption[]{Log of our \XMM observations of $\gamma^2$\,Vel. For a definition
of the phase see text of Section 2.
}
\label{obslog}
\begin{center}
\begin{tabular}{lcccc}
\hline\hline
\noalign{\smallskip}
 &  \multicolumn{1}{c}{high state}&
    \multicolumn{1}{c}{low state}&
    \multicolumn{1}{c}{high state}&
    \multicolumn{1}{c}{low state}\\
 &  \multicolumn{1}{c}{$\phi = 0.12$}&
    \multicolumn{1}{c}{$\phi = 0.34$}&
    \multicolumn{1}{c}{$\phi = 0.11$}&
    \multicolumn{1}{c}{$\phi = 0.37$}\\
 &  \multicolumn{1}{c}{8-11-2000}&
    \multicolumn{1}{c}{26-11-2000}&
    \multicolumn{1}{c}{14-4-2001}&
    \multicolumn{1}{c}{5-5-2001}\\
\noalign{\smallskip}
\hline\hline
\noalign{\smallskip}
instr. & exp. t.& exp. t. & exp. t. & exp. t.  \\
       & (hr) &  (hr) &  (hr) &  (hr)          \\
\noalign{\smallskip}\hline\noalign{\smallskip}
RGS1   & 4.29 & 5.80  & 8.72  & 16.81          \\
RGS2   & 4.51 & 5.80  & 8.72  & 16.81          \\
MOS1   & 3.76 & 5.74  & 8.56  & 16.64          \\
MOS2   & 3.83 & 5.77  & 8.56  & 16.64          \\
PN     & 2.78 & 4.16  & 7.83  & 15.14          \\
\noalign{\smallskip}
\hline \hline
\end{tabular}
\end{center}
\end{table}

Our observations were obtained at phases $\phi$\,$\simeq$\,0.11 and
$\phi$\,$\simeq$\,0.37.
The phase $\phi$ is calculated according
to the ephemeris of Schmutz et al. (1997). With this ephemeris,
periastron occurs at zero phase, the O-type component is in
front shortly afterwards at $\phi = 0.03$ and the WR is in front at
phase 0.61. The
observations were scheduled
according to the \xray light curve of Willis et al.  (1995).  The first
phase covers the short maximum, whereas at the second phase the \xray
flux is low.  The two observations of November 2000 were both terminated
prematurely because of strong solar radiation and, therefore, the exposure
times of the first two observations are considerably shorter than those of
April and May 2001.  Within the error bars the observations at
corresponding phases agree with each other. Here we use only the
second data set.

%%%%% FIGURE 2 %%%%%
\begin{figure}
\hbox{\hskip-0.2cm\psfig{figure=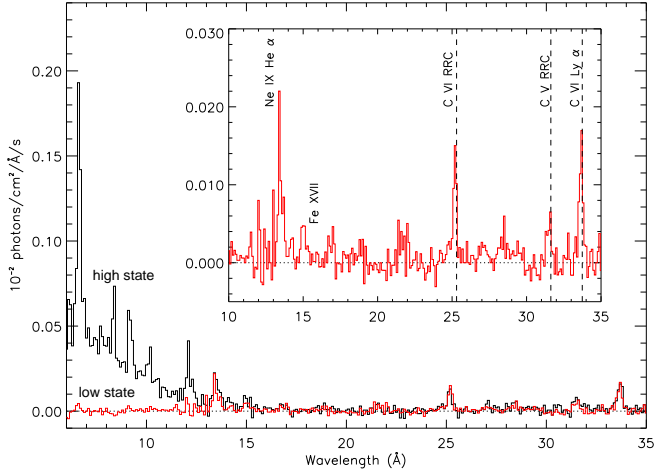,height=6.6cm,width=9.0cm,clip=}}
\caption{First order background subtracted {\sl XMM}-{\sc rgs} spectra of
$\gamma^2$\,Vel, corrected for effective area.  The spectra observed at
high ($\phi$\,=\,0.11) and low state ($\phi$\,=\,0.37) are plotted.  
The inset shows a blow-up of the $\phi$\,=\,0.37 observation.
        }
\label{figure2}
\end{figure}

%%%%% TABLE 2 %%%%%
%\begin{table}[t!]
\begin{table}
\caption{Measured line fluxes (in 10$^{-13}$\,erg\,cm$^{-2}$s$^{-1}$) at
Earth with  1\,$\sigma$ errors in parenthesis.} 
\label{lineflux}
\vskip-0.4cm
\begin{center}
\begin{tabular}{l@{\ }l@{\ }l@{\ }r@{\ }|l@{\ }l}
\hline \hline
%                &       &        &       &      &       \\[-1mm]
ion     &  ~E$_0$ &  E$_{\rm obs}$$^a$ &$\lambda_0$&  \multicolumn{2}{c} {flux} \cr
                & (keV) & (keV)  & (\AA) & high & low   \cr
%                &       &        &       &      &       \\[-1mm]
\hline\hline
                &       &        &       &      &       \\[-1mm] 
EPIC-MOS:       &       &        &       &      &       \\[-1mm]
Fe\,{\sc xxv}   & 6.701 & 6.669(10)   &  1.85 &  3.2(.3) & 3.9(.2)   \\
                & 6.637 &        &       &      &       \\
Ca\,{\sc xix}   & 3.903 & 3.919(22)   &  3.2  &  1.1(.2) & 0.66(.12)  \\
                & 3.861 &        &       &      &       \\
Ar\,{\sc xvii}  & 3.140 & 3.151(7)    &  4.0  &  2.0(.2)  & 0.41(.11)   \\
                & 3.104 &        &       &      &       \\
S\,{\sc xvi}    & 2.6229 &2.653(14)   &  4.7  &  1.2(.3)  & 0.28(.16)   \\
                & 2.6196 &        &       &      &       \\
S\,{\sc xv}     & 2.4607 & 2.466(4)   &  5.0  &  6.0(.3)  & 0.61(.11)   \\
                & 2.4473 &        &       &      &       \\
                & 2.4306 &        &       &      &       \\		
Si\,{\sc xiii}  & 2.1826 & 2.188(11)  &  5.7  &  1.2(.3)  & 0.05  \\
Si\,{\sc xiv}   & 2.0062 & 2.019(4)   &  6.2  &  3.2(.3)  & 0.05(.04)\\
                & 2.0043 &        &       &      &       \\
Si\,{\sc xiii}  & 1.8650 & 1.877(4)   &  6.7  &  5.5(.3)  & 0.30(.07)  \\
                & 1.8539 &        &       &      &       \\
                & 1.8395 & 1.833(9)   &       &  2.7(.2)  &       \\		
Mg\,{\sc xii}   & 1.4721 & 1.480(5)   &  8.42 & 1.6(.1)   & 0.11(.04)  \\
Mg\,{\sc xi}    & 1.3523 & 1.349(3)   &  9.17 & 2.2(.2)   & 0.30(.04)   \\
                & 1.3431 &            &  9.23 &      &        \\
                & 1.3312 &            &  9.31 &      &       \\
\hline			
RGS:            &        &            &       &      &          \\
Si\,{\sc xiv}   & 2.0062 & 2.00(1)    &  6.2  &  2.4(.9)  & --  \\
                & 2.0043 &            &       &      &       \\
Si\,{\sc xiii}  & 1.8650 & 1.859(3)   &  6.7  & 10.1(1.2) & 0.43(.27)  \\
                & 1.8539 &            &       &      &       \\
                & 1.8395 &            &       &      &       \\
Mg\,{\sc xii}   & 1.4721 & 1.474(2)   &  8.42 & 2.4(.4)   & --    \\			
Mg\,{\sc xi}    & 1.3523 & 1.3518(10) &  9.17 & 2.2(.4)   & 0.13(-)  \\
                & 1.3431 & 1.3336(20) &  9.23 & 1.6(.5)   & 0.09(.07)  \\
                & 1.3312 &            &  9.31 &           &       \\		
Ne\,{\sc x}     & 1.2109 & 1.2120(14) & 10.24 & 0.67(.18) & 0.14(.06)\\
Ne\,{\sc x}     & 1.0218 & 1.0258(16) & 12.13 & 1.19(.20) & 0.24(.08)  \\
Ne\,{\sc ix}    & 0.9220 & 0.9223(6)  & 13.44 & 0.59(.18) & 0.60(.11)  \\
                & 0.9148 & 0.9148(18) & 13.55 & 0.24(.16) & 0.16(.10)  \\
                & 0.9050 & 0.9038(9)  & 13.70 & 0.19(.14) & 0.34(.09)  \\
Fe\,{\sc xvii}  & 0.8258 & 0.8279(23) & 15.01 & 0.21(.08) & 0.15(.06)  \\		
O\,{\sc viii}   & 0.7746 & 0.7762(25) & 16.01 &  --       & 0.07(.05)  \\		
Fe\,{\sc xvii}  & 0.7392 & 0.7396(8)  & 16.78 &  0.14(.07)& 0.10(.03)  \\
Fe\,{\sc xvii}  & 0.7271 & 0.7286(-)  & 17.05 &  0.11(.07)& 0.07(.05)  \\
Fe\,{\sc xvii}  & 0.7251 & 0.7246(-)  & 17.10 &  --       & 0.12(.06)  \\
O\,{\sc viii}   & 0.6536 & 0.6543(20) & 18.97 &  --       & 0.17(.05)  \\
O\,{\sc vii}    & 0.5739 & 0.5742(31) & 21.60 &  --       & 0.12(.08) \\
O\,{\sc vii}    & 0.5686 & 0.5678(-)  & 21.80 &  --       & 0.03(-)\\
O\,{\sc vii}    & 0.5610 & 0.5600(17) & 22.10 &  --       & 0.13(.06)  \\
N\,{\sc vii}    & 0.5003 & 0.5009(10) & 24.78 &  --       & 0.24(.13)\\
C\,{\sc vi} RRC & 0.4900 & 0.4911(3)  & 25.30 & 0.30(.09) & 0.40(.07)   \\
C\,{\sc vi}     & 0.4356 & 0.4357(10) & 28.47 & 0.16(.11) & 0.06(.03)\\
N\,{\sc vi}     & 0.4198 & 0.4188(3)  & 29.53 & 0.14(.08) & 0.07(.04)\\
C\,{\sc v}  RRC & 0.3921 & 0.3927(8)  & 31.60 & 0.18(.13) & 0.12(.06)   \\
C\,{\sc vi}     & 0.3675 & 0.3679(4)  & 33.74 & 0.31(.08) & 0.35(.09)  \\[-1mm]
                &       &        &       &      &       \\
\hline\hline
\end{tabular}
\end{center}
\begin{flushleft}
{
\begin{description}
\item $a$: From high (or low) state for $E_{\rm obs} >$ (or $<$)~1~keV, resp. 
\end{description}
}
\end{flushleft}
\end{table}

The spectra were obtained with the \XMM {\sc e}uropean {\sc p}hoton {\sc
i}maging {\sc c}ameras ({\sc epic}), {\sc mos} and {\sc pn}, and with the
high-resolution {\sc r}eflection {\sc g}rating {\sc s}pectrometers ({\sc
rgs}).  The {\sc rgs} have an energy coverage from 0.35 to 2.5\,keV (5 to 37\,\AA), and
the {\sc epic} can exploit the full XMM range from 0.15 to 15\,keV (0.8 to 83\,\AA).

The data have been reduced with standard procedures using the \XMM {\sc
s}cience {\sc a}nalysis {\sc s}ystem ({\sc sas}), with the available
calibration data.  The {\sc epic} response matrices from the {\sc
epic} instrument team have been used to fit the {\sc ccd} spectra.  The
{\sc rgs1} and {\sc rgs2} spectra have been co-added.  No correction for
interstellar absorption has been applied.

\section{Spectral features and variability}

In this section we highlight and discuss selected features of the observed
X-ray spectra.  We apply simple analysis techniques in order to get
some indications about the underlying excitation, ionization and absorption
mechanisms.

%%%%% FIGURE 3 %%%%%
\begin{figure}
\hbox{\hskip-0.2cm\psfig{figure=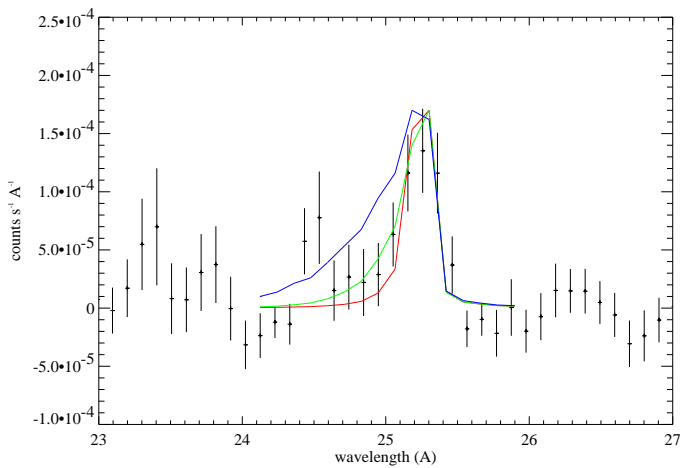,width=9.0cm,clip=}}
\caption{Observed C\,{\sc vi} recombination continuum with the theoretically
calculated energy distribution folded with the
{\sl XMM}-{\sc rgs} response function. The profiles correspond to electron
temperatures $T_{\rm e}$\,=\,80\,000\,K, 40\,000\,K, and
20\,000\,K (from  wide to narrow).
        }
\label{figure3}
\end{figure}

\subsection{The emission lines\label{emis}}

Our high-resolution ({\sc rgs}) spectra are shown in Fig.\,\ref{figure2}.  Line
identifications and measured line fluxes are listed in
Tab.~\ref{lineflux}.  The line fluxes given by Skinner et al. (2001) and
taken at phase 0.08 lie mostly between our high and low line state
line fluxes.  This is consistent with the steep increase in X-ray intensity
at early phases.

\subsubsection{Carbon recombination features}

Surprisingly we find among these emissions the radiative recombination
continua (RRC) of \Cvi and \Cv at 25.3\,\AA\ and 31.6\,\AA,
respectively
(Fig.\,\ref{figure2}).
Also the \Cvi\,Ly$\alpha$ line is
clearly detected.  The intensity of this line with respect to the RRC
is in agreement with it being formed by recombination.  The shape
of the high energy tail of the RRC is
a direct measure of the temperature of the recombining electrons.  In
Fig.\,\ref{figure3} we compare the observed C\,{\sc vi} RRC with the
theoretically calculated spectral distribution for different electron
temperatures convolved with the {\sc rgs} response curve.  The 
temperatures between about 60\,000\,K and 20\,000\,K agree with
the observed energy distribution.  This temperature is too low for
collisional ionization and implies that radiation may be the dominating
ionization process. A possible radiation source could be the \xray emission 
from the wind-wind collision zone.

\subsubsection{Line width}

The \Ovii and \Oviii emission lines in the low state spectrum of
$\gamma^2$\,Vel, and the \Cvi\,Ly$\alpha$ (33.74\,\AA) line in the
high and low state spectra, show some measurable
broadening.  Fitting the \Cvi line
with an instrumental profile results in
{\sc fwhm}\,=\,1.8\,$\pm$\,0.8\,$\times$10$^{-3}$\,keV for both, low  
and high state, corresponding
to an expansion velocity of the order of 1300\,km\,s$^{-1}$.  Such
velocities are comparable to those found in the {\sl Chandra}-{\sc hetgs}
spectrum of $\gamma^2$\,Vel for He- and H-like lines of Mg, Si, and S
(Skinner et al. 2001), and are close to the terminal wind velocity
$v_{\infty}$\,=\,1450\,km\,s$^{-1}$ of the WC star, as found from infrared
\Hei lines by Eenens \& Williams (1994).

\subsubsection{The Ne\,{\sc ix} lines}

%%%%% FIGURE 4 %%%%%
\begin{figure}
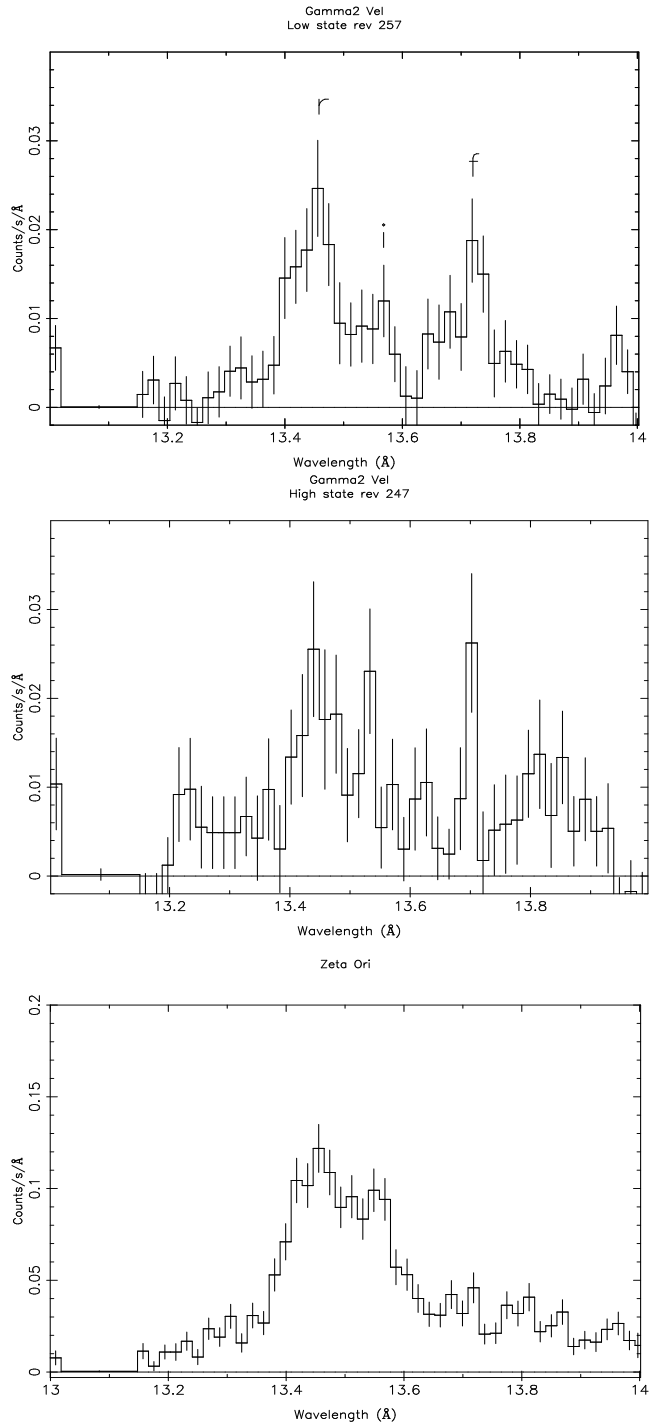

\vbox{
\hbox{\psfig{figure=35fig4to.ps,angle=-90,height=6.2cm,width=8.5cm,clip=}}
\hbox{\psfig{figure=35fig4mi.ps,angle=-90,height=6.2cm,width=8.5cm,clip=}}
\vskip 0.2cm
\hbox{\psfig{figure=35fig4bo.ps,angle=-90,height=6.2cm,width=8.5cm,clip=}}
\caption{\Neix triplet consisting of a resonance line ($r$), an
intercombination line ($i$), and a forbidden line ($f$) for 
$\gamma^2$\,Vel ({\it top and center}) and for the O9.7Ib star $\zeta$\,Ori ({\it bottom}),
observed with {\sl XMM}-{\sc rgs}.  Note the difference in the forbidden
line. }
\label{neix.fig}
}
\end{figure}

Figure~\ref{neix.fig} shows the resonance ($r$), intercombination ($i$)
and forbidden ($f$) components of the He-like \Neix triplet
of $\gamma^2$\,Vel (WC8+O7.5III) and compares it with with the same
lines observed by {\sl XMM-Newton}-{\sc rgs} in the O-type system
$\zeta$\,Ori
(O9.7Ib\,+\,O8-9V\,+\,B2III, see Hummel et al. 2000).

The forbidden line in $\gamma^2$\,Vel 
is much stronger than in O stars. In  $\gamma^2$\,Vel the ratio 
$R$ =$f$/$i$ at low
state is consistent with  $R_{0}$=3.1, the value expected when excitations
from the upper level of the forbidden line to the upper levels of the
intercombination line are
negligible. This is in contrast to observations of single O-type stars
(Schulz et al 2000, Kahn et al 2001, Waldron \& Cassinelli 2001,
Cassinelli et al 2001, Miller et al 2002, Raasen et al 2004), where R is
smaller. In these O stars, the X-ray emitting plasma is formed relatively
close to the star, and UV photoexcitation from the upper level of 
the forbidden to
the upper levels of the intercombination line is responsible for the 
observed ratio. The high $R$
ratio observed in gamma Vel implies that the Ne IX lines are formed far
away from the O star, where the UV flux is low.

Our observation of the \Neix triplet at high state (Fig.~\ref{neix.fig}, central panel)
is much noisier. Although the
total flux is approximately the same, the exposure time is about half as
long, and the statistics are correspondingly worse. The $R$ ratio appears
to be closer to about 1 in this case, as compared to $R_{0}$=3.1, but it
is not clear whether this is statistically significant. Chandra HETG
obervations of gamma Vel at a similar phase find $R > \frac{2}{3} R_0$ for
\Neix (Skinner et al. 2001).

A lower limit to the distance from the O star at which the \Neix lines 
are formed can be
calculated using the formalism of Blumenthal, Drake \& Tucker (1972).
The dependence of $R/R_0$ on radius can be written
$R/R_0=1/(1+\psi/\psi_c)$ where $\psi_c=7.73\times10^3\rm{s^{-1}}$,
$\psi=\psi_*(1-\sqrt(1-(\frac{R_*}{r})^2))$, and $\psi_*=10^6\,\rm{s^{-1}}$
is the photoexcitation
rate at 1278 \AA\ at the photosphere which is calculated with 
a TLUSTY model (see Lanz \& Hubeny, 2003) with
log g=3.75 and T=35000 . Taking $R > \frac{2}{3} R_0$,
we find $r > 11 R_*$, with $R_*$ the photospheric radius of the O
star. 
The biggest uncertainties are the photospheric
UV flux, which is a strong function of wavelength due to the many
aborption lines near this wavelength, and the f/i ratio itself.

It is clear that the Ne IX emission is not coming from anywhere near the
O star, and this essentially rules out the possibility that it arises due
to intrinsic X-ray emission from the stellar wind of the O star.

We finally note that the deduced emission measure of the \Neix lines
is surprisingly small (Dumm et
al.  2003) compared to the emission measure of the single O4I(n)f star
$\zeta$\,Pup (Kahn et al. 2001).

\subsection{Absorption at low state \label{abs}}

\begin{figure}
\hbox{\hskip-0.2cm\psfig{figure=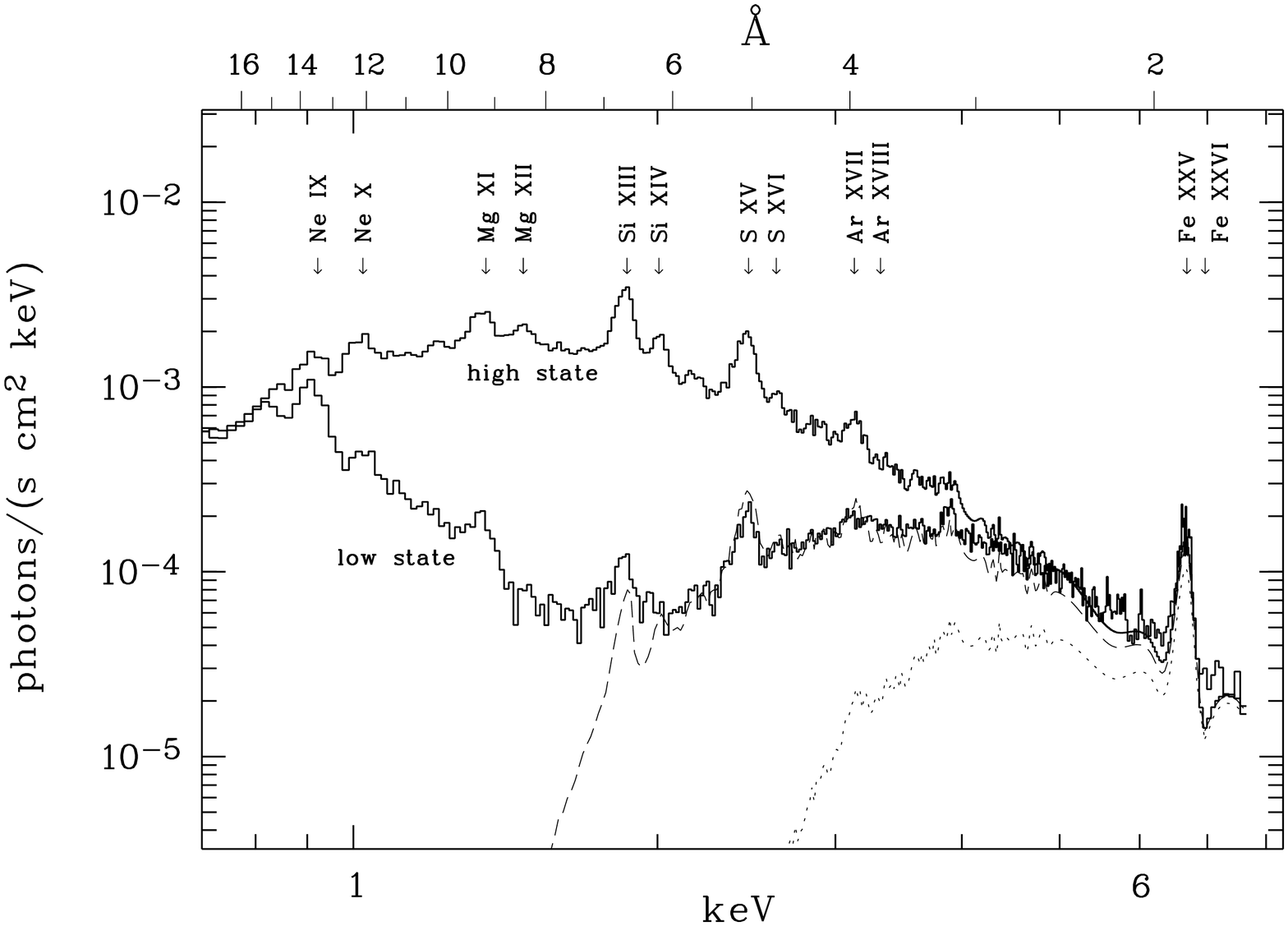,height=7cm,width=9.5cm,clip=}}
\caption{
{\sl XMM}-{\sc epic-mos} data of $\gamma^2$\,Vel at low and
high state (full lines),
observed in 2001.  The dotted line is the high state
spectrum absorbed by the column density predicted by a smooth
WR wind model.  The dashed line
is the same with a column density reduced by a factor of 4 (see text).
}
\label{figure5}
\end{figure}

Fig.\,\ref{figure5} shows our {\sl XMM}-{\sc epic} spectra at the low and
high states.  At very low and very high energies the spectra are identical,
but at intermediate energies the low state spectrum shows a deep
depression.  At $\sim$\,2\,keV the photon flux is reduced by
more than an order of magnitude. At low state, a flux deficiency is detected between
about 1 and 4\,keV.
The emission lines also show a different absorption behaviour. Some lines are
heavily absorbed while others are not:
\vspace{-3mm}
\begin{itemize}

\item{The absorbed component:}
At phase $\phi$\,=\,0.37, the hard emission lines are all reduced by a
large factor with the exception of the \Fexxv complex  at 1.85\,\AA.
(Tab.~\ref{lineflux}). The lines of the
highly ionized species of Ar, S, Si, and Mg which are very strong at high
state are barely detectable at low
state.  We therefore expect
that these lines originate in the deeply embedded wind-wind collision zone.

\item{The unabsorbed component:}
There is no measurable line flux variations longward of 13\,\AA.  We
detect here lines from the
Fe\,{\sc xvii}, Ne\,{\sc ix}, O\,{\sc viii}, O\,{\sc vii}, and C\,{\sc vi}
species. The radiative recombination continua (RRC) of \Cvi\,25.3\,\AA\ and
\Cv\,at 31.6\,\AA\ are also not absorbed.  
These practically unchanging emissions are formed further
out in the binary system, such that any wind material between us and the emission region is
transparent. We note that the unabsorbed emissions originate from
two different mechanisms, i.e., from collisional excitation and
recombination.

\end{itemize}

%%%%% FIGURE 5 %%%%%

\section{Absorption by the Wolf-Rayet wind\label{wrclumps}}

\subsection{The absorption column and mass loss rate}

To interpret the absorption at phase 0.37, we adopt the view of Willis et
al. (1995) and Stevens et al. (1996), in that we assume that the high-energy
X-rays emerge from the wind-wind collision zone.  These X-rays are
attenuated by the surrounding material of the WR wind, except for the phase
where we view the collision zone through the cavity behind the O-type
component (see Fig.\,\ref{figure1}). The attenuation of the X-rays during low
phase is a direct measure of the column density in the WR wind. The contribution
of the O star wind and the shock zone to the total absorption column is small 
compared to the WR wind and we neglect it.

We calculate the expected absorption at phase $\phi=0.37$ by using a smooth
WR wind model that was especially developed for $\gamma^2$\,Vel (De\,Marco et
al. 2000, {\sc isa} model). We approximate
the position of the X-ray source with the O star itself.  The location of
the stagnation point is given by the wind momentum balance

\be
\lambda = \sqrt{\frac{\Mdot_{\rm WR}\,v_{\rm WR}}{\Mdot_{\rm O} v_{\rm O}}}
= 5.9 \ee

This high $\lambda$ ratio, obtained with the mass loss data of
De\,Marco \& Schmutz (1999) and De\,Marco et al. (2000) implies that the
collision zone is indeed very close to the O-type component.  

The low
state spectrum is calculated by using the high state spectrum as
input and passing it through the absorption column from the position
of the O star to
the observer.
We implicitely assume that the \xray emission does
not vary with phase.  This is not necessarily correct because theoretically,
we might 
expect a \xray luminosity that is inversely proportional to the instantaneous
binary separation (Usov 1992).
Since, however, we do not see any significant 
difference between high and low
state above 4.5\,keV ($<$ 2.7\,\AA), we believe that the assumption of
a constant \xray emission 
is reasonably well justified.

The WR atmosphere model includes helium,
carbon and oxygen abundances and their ionization states.  Neon 
is assumed to be doubly ionized and its abundance is set to 
Ne/He\,=\,4$\times$10$^{-3}$ (by
number) as determined from {\sl ISO}-{\sc sws} spectroscopy (Dessart et al.
2000).  The magnesium and silicon
abundances are set to $1/10$th of the neon abundance, and for the sulfur
abundance we adopt $1/20$th of the neon abundance.  Errors in these
abundances only have a marginal effect on the overall appearance of the
emerging spectrum. Analytical fits for the partial photo-ionization
cross sections given by Verner \& Yakovlev (1995) are used.

In Fig.\,\ref{figure5} we indicate by the dotted line the
attenuation at phase $\phi$\,=\,0.37 predicted by this
WR wind model.  The 
absorption by this WR wind is substantially
too high. If we
treat the column density as a free parameter and adjust it to fit the
observed attenuation, we derive a column density of
$N_{\rm tot}$=5$\times$10$^{21}$\,cm$^{-2}$. This is a factor 4 smaller
than what it
would be in the WR wind model.

From the
observed colum density we can calculate a distance-independent mass
loss rate of 8$\times$10$^{-6}$\,\Msolar/yr for the WC8 star.
The dominating error source
in this value is uncertainties in the chemical
composition, particularly in the carbon and
oxygen abundances because these elements dominate the opacity in the wind.
A further but probably minor factor is the assumption of spherical
symmetry in the extended WR atmosphere.

\subsection{Wolf-Rayet wind clumping}

The fact that the mass loss rate of the WC8 star determined through \xray absorption
is a factor of four less than what is predicted by a homogeneous
 atmospheric model deduced from spectral line fits
is easily interpreted in terms of Wolf-Rayet wind clumping.

Clumping of a stellar wind
can be described in a simple way by the clumping factor $f$. This factor
defines by how
much the density $n_{clumped}$ is enhanced in a clump with respect to
the 
smooth density $n_{smooth}$

\be
n_{clumped} = f\times n_{smooth}
\ee

The interclump volume is assumed to be devoid of material.
In order to keep the total amount of material the same in the clumped
and the smooth model, the clumps only
fill a volume that is $1/f$ of the volume of the smooth
model. The volume filling factor is thus the inverse of the clumping factor.

\be
V_{clumped} = \frac{V_{smooth}}{f}
\ee

WR model atmospheres resulting from emission line fits can not
easily disentangle the effects of mass loss and clumping.
Locally, the intensity of a WR emission line is determined by $n^2$,
with $n$ the local particle density.
The total line
emission in a smooth model is 

\be
I \propto \int_{V_{smooth}} n_{smooth}^2\, dV
            \propto \Mdot^2
\ee

$\Mdot$ is the mass loss rate that follows from fitting the model
to the WR emission lines. 

In the case of a clumped wind the line intensity becomes 

\bea
I &\propto& \int_{V_{clumped}} n_{clumped}^2\, dV    \\
  &=& \int_{V_{clumped}} f^2\,n_{smooth}^2\, dV  \\
  &=& f^2 \int_{V_{smooth}} \frac{n_{smooth}^2}{f} \,dV  \\
  &\propto& f\times\Mdot^2
\eea

The same line intensity can thus be produced by various combinations of
the mass loss rate and the clumping factor.

Since $f$ is larger than 1, a clumped wind always produces stronger
emission lines than a smooth wind with the same amount of matter in it.

In the previous section we found from the \xray absorption that the
mass loss rate of the WR star in  $\gamma^2$\,Vel is only
one fourth of the rate deduced from a smooth Wolf-Rayet
wind model. In order to still fit the emission lines the clumping
factor has to be

\be
f = 16
\ee

In this scheme, the volume filled by the clumps is thus only 6\,\%.

We note that this clumping factor follows from our mass loss
rate as determined from the observed \xray absorption in conjunction with
a Wolf-Rayet model atmosphere that is based on UV, optical and IR emission line
fitting. The mass loss rate of the smooth model depends on the
adopted distance $d$: $\Mdot\,\propto\,d^{3/2}$. The model we use
is for $d$=258\,pc. If we adopt instead of the {\sl Hipparcos} distance the
older value of $d$=450\,pc, the mass loss of the smooth model would become 2.3
times larger. The mass loss rate from the \xray absorption, on the other hand,
is not affected by the distance. The \xray deduced mass loss rate would in that
case be about 9 times smaller than the one from the smooth model. The
clumping would have to be much more pronounced with a clumping factor
as high as $\sim$\,80.

\section{Synthesizing the \xray emission}

We now turn to the interpretation of  the \xray {\it emission}.
In Sect.\,3 we have already seen that the
\xray spectrum consists of different components.  In
order to further deepen the analysis we now develop synthesized spectra
to match the full observed data at both low and high state.

\subsection{The basic models}

The spectra as discussed above obviously require synthetic
models that should include hot thermal sources (predominantly for 
the emission between 1--10~keV) but also a recombination model
(for the long-wavelength portion). The basic model components we use
for the description of all spectra are included in the
{\sc SPEX} software (Kaastra et al. 1996). For the hot thermal
sources, we use optically thin plasma models in collisional ionization 
equilibrium ({\sc cie} models) as developed by Mewe et al. (1985, 1995).
The underlying {\sc mekal} data base is given as an extended list of 
fluxes of more than 5400 spectral lines.
\footnote{http://www.sron.nl/divisions/hea/spex/version1.10/line/}

%%%%% TABLE 3 %%%%%
\begin{table}
\caption{Fixed and fitted elemental abundances of the \xray
emitting material in $\gamma^2$\,Vel. The abundances are given relative
to He in units of the solar ratio. The solar abundances are from Anders \& Grevesse
(1989), or from
Grevesse and Sauval (1998, 1999) for Fe. Abundances given in 
italics were
allowed to vary in the fit and the others were kept fixed.}
\label{abu.tab}
\begin{center}
\begin{tabular}{   l@{\ }|c@{\ }c@{\ }}
\hline\hline
element &\multicolumn{2}{c}{ionization}  \\
        & collisional       & photo-     \\
\hline
H       & 0.5               & 0          \\
He      & 1                 & 1          \\
C       & 14                & 27         \\
N       & 0.5               & 0          \\
O       & 1.6               & {\it 2.2}  \\
Ne      & {\it 2.5\,(.5)}   & 2.4        \\
Mg      & {\it 1.2\,(.1)}   & 1          \\
Si      & {\it 1.07\,(.08)} & 1          \\
S       & {\it 1.00\,(.08)} & 1          \\
Ar      & {\it 1.0\,(.15)}  & 1          \\
Ca      & {\it 1.2\,(.2)}   & 1          \\
Fe      & {\it 1.0\,(.1)}   & 1          \\
Ni      & {\it 2.4}         & 1          \\
\hline\hline
\end{tabular}
\end{center}
\end{table}

The long-wavelength part of the spectrum is interpreted with a model
in which lines are formed by pure radiative recombination, e.g., 
C~{\sc vi} Ly$\alpha$, $\beta$, $\gamma$ lines at 33.74~\AA, 28.47~\AA\ and
26.99~\AA, and the O~{\sc vii} lines at 21.6--22.1~\AA. The shape of the
{\sc rrc} at 25.3~\AA\ constrains the temperature of the emitting
gas to $T$\,$\simeq$\,38\,000\,$\pm$\,7\,500\,K
(cf. Fig.\,\ref{figure3}). 
The {\sc rrc} of C~{\sc v} at
31.6~\AA\ is also fitted by this model. Methodologically, this
component is described by a temperature-jump model in {\sc spex}
for which we assume a relatively high starting temperature (e.g., 1~keV) and a steep
temperature drop (of the order of a few $10^4$~K), leaving the plasma
in a purely recombining state.

%%%%% TABLE 4 %%%%%
\begin{table*}[t!]
\caption{Best 4 component fit for combined {\sl XMM} spectra of $\gamma^2$\,Vel at the two phases.
Components 1 to 3 are collisionally ionized and component 4 is
photo-ionized.  $L_{\rm xi}$ is the unabsorbed \xray luminosity in the range
0.4\,-\,10\,keV for each component.  1\,$\sigma$ uncertainties are given in
parentheses.  The {\sl Hipparcos} distance to $\gamma^2$\,Vel of
$d$\,=\,258\,pc is adopted.  For comparison, a two-component fit to similar
data for $\zeta$\,Ori (O9.7Ib) is given.
        }
\label{fitres.tab}
\begin{center}
\begin{tabular}{ l@{\ }               || l@{\ }      l@{\ }      l@{\ }       l@{\ }     | l@{\ }        l@{\ }             || l@{\ }       l@{\ }      l@{\ }      l@{\ }     | l@{\ }         l@{\ }  }
\hline \hline
                                       &           &           &            &             &             &                     &            &           &           &            &              &                    \\[-1mm]
                                       & \multicolumn{4}{c|}{$\gamma^2$\,Vel}             & \multicolumn{2}{c||}{$\zeta$\,Ori}& \multicolumn{4}{c|}{$\gamma^2$\,Vel}            & \multicolumn{2}{|c}{$\zeta$\,Ori} \\
                                       & \multicolumn{4}{c|}{$\phi$\,=\,0.11, high state} &             &                     & \multicolumn{4}{c|}{$\phi$\,=\,0.37, low state} &              &                    \\[-1mm]
                                       &           &           &            &             &             &                     &            &           &           &            &              &                    \\[-1mm]
\hline\hline
                                       &           &           &            &            &             &                     &            &           &           &            &              &                    \\[-1mm]
component i                            & 1         & 2         & 3          & 4          & 1           & 2                   & 1          & 2         & 3         & 4          & 1            & 2                  \\[+0.3mm]
                                       &           &           &            &            &             &                     &            &           &           &            &              &                    \\[-1mm]
\hline
                                       &           &           &            &            &             &                     &            &           &           &            &              &                    \\[-1mm]
$N_{\rm Hi}$[10$^{22}$cm$^{-2}$]$^a$   & 3.3(.4)   & 1.87(.09) & 1.04(.05)  & 0.011      & 1.1         & 1.1                 &  14.6(.6)  & 7.1(.5)   & 0.84(.02) & $\la$0.041 & 6.0          & 6.0                \\[+0.3mm]
$N_{\rm Xi}$$^b$                       &           &           &            &            &             &                     & 5.1(0.3)   & 2.4(.2)   &           &            &              &                    \\[+0.3mm]
$T_{\rm i}$ [keV]                      & 1.65(.07) & 0.66(.03) & 0.25       & 0.0033$^c$ & 0.57        & 0.20                & 1.53(.04)  & 0.67(.06) & 0.25(.07) & 0.0033$^c$ & 0.57         & 0.20               \\[+0.3mm]
$EM_{\rm i}$[10$^{55}$cm$^{-3}]$       & 2.09(.18) & 3.9(.3)   & 0.48       & 0.0031$^c$ & 0.14        & 0.40                & 4.0(.3)    & 1.6(.5)   & 0.48(.05) & 0.0031$^c$ & 0.14         & 0.40               \\[+0.3mm]
$L_{\rm xi}$[10$^{32}$\,erg\,s$^{-1}$] & 3.1       & 10.4      & 1.3        & 0.012      & 0.41        & 0.54                & 6.1        & 4.2       & 1.3       & 0.012      & 0.41         & 0.54               \\[+0.3mm]
$L_{\rm x}=\sum_{\rm i} L_{\rm xi}$    &           & 14.8      &            &            &             &                     &            & 11.6      &           &            &              &                    \\[+0.3mm]
\gired                                 &           & 2263/1602 &            &            &             &                     &            & 2232/1602 &           &            &              &                    \\
                                       &           &           &            &            &             &                     &            &           &           &            &              &                    \\[-1mm]
\hline\hline
\end{tabular}
\end{center}
\begin{flushleft}
{
\begin{description}
\item Notes:
\item $a$: Column density derived for standard absorption model (with
solar abundances) from Morrison \& McCammon (1983) for all components. In all fits an
additional value of 0.008$\times$10$^{22}$\,cm$^{-2}$ was assumed for the interstellar absorption.  The
results indicate that most of the absorption must be due to circumstellar
material.
\item $b$: Column density with a
warm absorber model in {\sc spex} with temperature 50\,000\,K and
abundances according to those in the wind of a WC star (see Table 3, 2nd column).
Because here the H abundance is zero
we use the subscript X referring to the reference element X (here He). Though the
values for $N_{\rm X}$ are smaller than the corresponding values for $N_{\rm H}$ in the previous fit,
the overall absorption is still the same because of the larger contribution
from other elements (e.g. C). The fit is significantly the same because \gired\,=\,2215/1606.
\item $c$: Temperature and emission measure determined by a fit
to the \Cvi {\sc rrc} feature at 25.3 \AA\ in the low state (errors are
0.0006\,keV and 0.0008$\times$10$^{55}$\,cm$^{-3}$, respectively).  For the high state we assume the same values.
\end{description}
}
\end{flushleft}
\end{table*}

In principle, each of the model components could have its own set of
abundances. Fortunately, however, we have relatively good a priori knowledge
of the composition of the two stellar winds, and we make the following plausible
assumptions.

For the recombining component, we assume a pure WC star plasma, because this 
is suggested by the dominance of the C lines. The abundances for
the WC wind were taken from van der Hucht et al. (1986), De Marco 
et al. (2000), and Dessart et al. (2000). The abundances in
Tab.\,\ref{abu.tab} are given
relative to the solar photospheric values. The solar abundances are
from Anders \& Grevesse (1989) except for Fe, for which we
use log\,$A_{\rm Fe}$\,=\,7.50 (see
Grevesse \& Sauval 1998 and 1999) instead of 7.67.
The WC abundances  largely deviate from 
solar composition in particular for He, C, N, O, and Ne. In the fit we
keep all abundances fixed except for O.

The elemental abundances of the hot, collisionally ionized plasma lie somewhere
between the abundances of the WR star and those of the O star. 
We adopt a mixture of material with 50\% WC and 50\% solar composition.
This is justified because the difference
between solar and WC abundance patterns is mainly in the light elements
up to neon.
Heavier elements are only slightly or not at all processed in WC stars.
 The
abundances of all elements with observable emission lines are left to vary
while those with no emission lines are kept fixed (Tab.~\ref{abu.tab}).

\subsection{Model strategy} 

We developed an acceptable model along the following line. First, during low
state, there is a hard component isolated from the longer wavelengths by the
absorption trough. This component does not vary between low and high state
but it is subject to variable absorption. We found that two {\sc cie}  components
are required to describe the range of temperatures forming lines from Mg~{\sc xi}
to Fe~{\sc xxv} plus the hot continuum in the high state.

To first order, the {\it low-state} spectrum
longward of  about 8~\AA\ or below about 1.5~keV is uncontaminated by
the hard component. Judged
from the flux in the Ne~{\sc IX} line, it is also approximately constant 
between low and high state, with {\it constant} absorption. This component alone 
therefore describes the low-state {\sc cie} plasma at long wavelengths.

In a first iteration we, therefore, used three {\sc cie} models, 
each with constant emission measure, of which only the two hotter ones are
subject to variable absorption between low and high state, and all 
are composed of a mixture of WR and O star material as indicated in
Table~\ref{abu.tab}.
A separate fourth component, again constant for low and high state, is required for
the recombining portion longward of $\approx$15~\AA, with a pure WC composition.

Using this setup, the interpretation becomes essentially one of the
absorption of the various components. During high state, the absorption is obviously weak for all components, which
is little surprising as the radiation principally escapes through the
low-density cone behind the O star. We therefore assume solar composition for
the absorption components during high state.
The deep absorption of the hotter material in low state, on the other hand, is 
thought to be due to the WR wind when it moves into the line of sight toward the
hot shock region. Its composition should therefore be identical to the WR wind, 
which we assumed in our calculations although we also tested absorption by a wind with
solar composition e.g., from the O star. The results are similar although
the definition of the column density necessarily varies (Table 3).

We tested in a first iteration whether this setup produces meaningful results.
To this end, we fitted the
low-state spectrum longward of $\approx$\,8\AA\ (below 1.5~keV) with the
coolest {\sc cie} model, and simultaneously the complete high-state spectrum
with this same model plus the two hotter {\sc cie} components. The recombining
portion was not included, and the isolated, absorbed high-energy portion of the
low state was also not considered. Each of the three {\sc cie} components 
was subject to a separate but constant absorption. Since the abundance mix
should not be too far from solar (apart from C, N, and O which are not
varied
for the {\sc cie} components since the relevant lines are formed longward of
15\,\AA), we determined them, in a first step - and deviating from our
final choice described in Sect.~5.1 -  as free parameters
without using  a priori assumptions. Spectra from both RGS
and both MOS detectors were fitted simultaneously with this composite model.

The three resulting temperatures were found to be approximately
$0.23, 0.65,$
and $1.8$~keV, with emission measures in the proportion of  1.0:4.5:2.0.
The
most interesting result was increasing absorption columns,
$N_H = 8.8\times 10^{21}$\,cm$^{-2}$,
$N_H = 1.7\times 10^{22}$\,cm$^{-2}$, and
$N_H = 2.5\times 10^{22}$\,cm$^{-2}$, respectively. This suggests that
the hot
material is more deeply embedded in the WR wind than the cool material.

Most abundances were found within a factor of two around solar photospheric
values (C, N, O not included), making the assumption of a 
mixture between O and WR star material as suggested in Table~\ref{abu.tab}
plausible. With
a $\chi^2$ value of 1.55, the overall spectral fit was acceptable although 
deficiencies, in particular in the region around 1 keV, were still visible.

%%%%% FIGURE 6 %%%%%
\begin{figure}
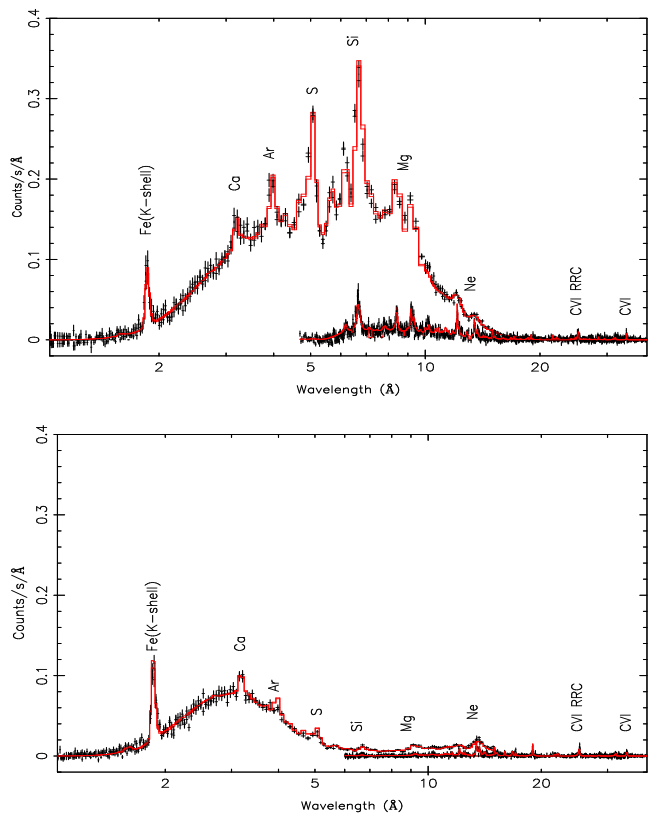

\hbox{\psfig{figure=35fig6to.ps,angle=-90,height=5.5cm,width=8.5cm,clip=}}
\hbox{\psfig{figure=35fig6bo.ps,angle=-90,height=5.5cm,width=8.5cm,clip=}}
\caption{The data points show the first-order background-subtracted 
{\sl XMM}-{\sc epic-mos} and
-{\sc rgs} spectra  at phases $\phi$\,=\,0.11 (high
state; {\it upper panel}) and $\phi$\,=\,0.37 (low state; {\it lower
panel}).  Error bars correspond to 1\,$\sigma$ statistical errors including
the effect of background.  The plotted curves show our best-fit model.
        }
\label{fitspec.fig}
\end{figure}

\subsection{Spectral synthesis} 

With this plausibility analysis, we proceeded with a more general model
that, first,
also includes self-consistently the variable WR wind absorption of the
hot components, second, independently fits
the recombination component with WR abundances (Tab. 3). The coolest {\sc cie}
component was fitted in
the low state and was then fixed for the high state (this
is borne out
by the constancy of the Ne~{\sc ix} line flux in the spectra, see, e.g.,
Fig. 2).
The abundances are again fitted  except for H, C, N, and O for which
we use the predetermined 
mixture from the O and WR star winds. The other elemental
abundances, in fact, should
be easy to determine since the spectra contain isolated and well developed
line features for Ne, Mg, Si, S, Ar, and Fe in the high state
(Fig.\,5, Tab.\,2). The abundance results are summarized in
Tab.\,\ref{abu.tab}. Most abundances are compatible with
a solar composition with the notable exception of Neon,
which turns out to be 2.5 times solar.
Of the abundances that we fit,
Neon is the one that is most sensitive to discriminate between
a solar and WC composition.
Although clearly enhanced, our \xray determined
Neon abundance is lower than the factor of 6-8 times solar that is measured
in the WC8 
wind with
ISO spectroscopy of WR wind lines (Dessart et al. 2000).
We therefore clearly detect a contribution from the WR star in the shocked
material
but the Neon abundance is still lower than what would be expected if it
were purely
composed of WC material. This indicates that the shocked material is
indeed a mixture
of WC and O star material. Our measured Neon abundance also justifies 'a
 posteriori' our choice of abundances as a mixture of half WR material
 and half
 O star material for those elements that are not observable through their
 \xray emission lines.

Table~\ref{fitres.tab} summarizes the results of the fitting procedure
for the various emission components.
It shows a gratifying agreement with our previous
plausibility analysis. The
temperatures are essentially the same, and the absorption column densities in 
high state increase with increasing temperature of the component. The total
emission measure of the two hotter components stays constant although the
distribution  between the two components differs, which we suggest is
a numerical artifact since a slight uncertainty in the absorption columns,
in particular during low state, requires a large correction in the emission
measures. We note that with the exception of the \xray luminosities
$L_{\rm x}$,
the values in Table~\ref{fitres.tab} do not depend on the distance.

\begin{figure*}
\vbox{
\hbox{
\psfig{figure=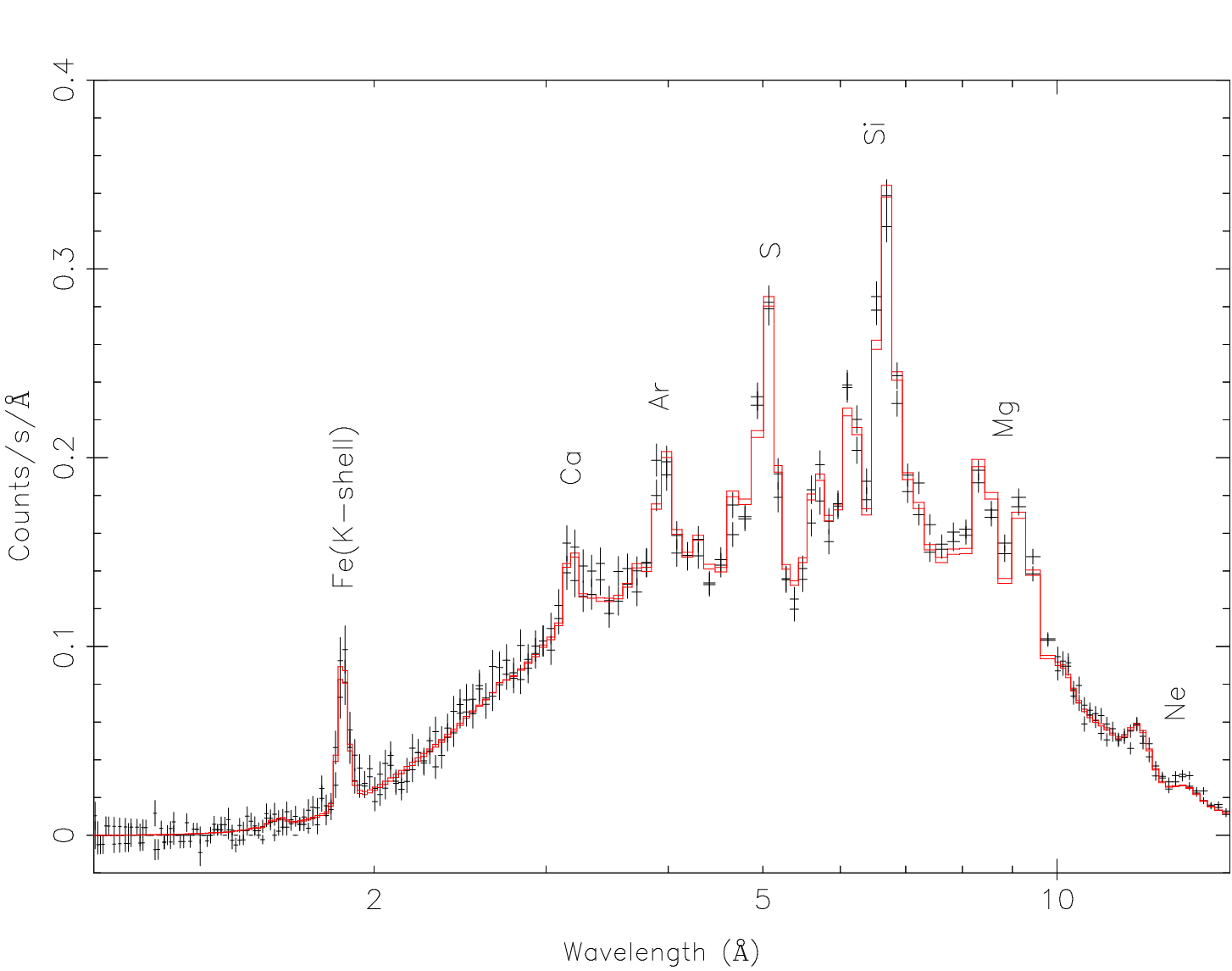,angle=-90,height=5cm,width=8.5cm}
\psfig{figure=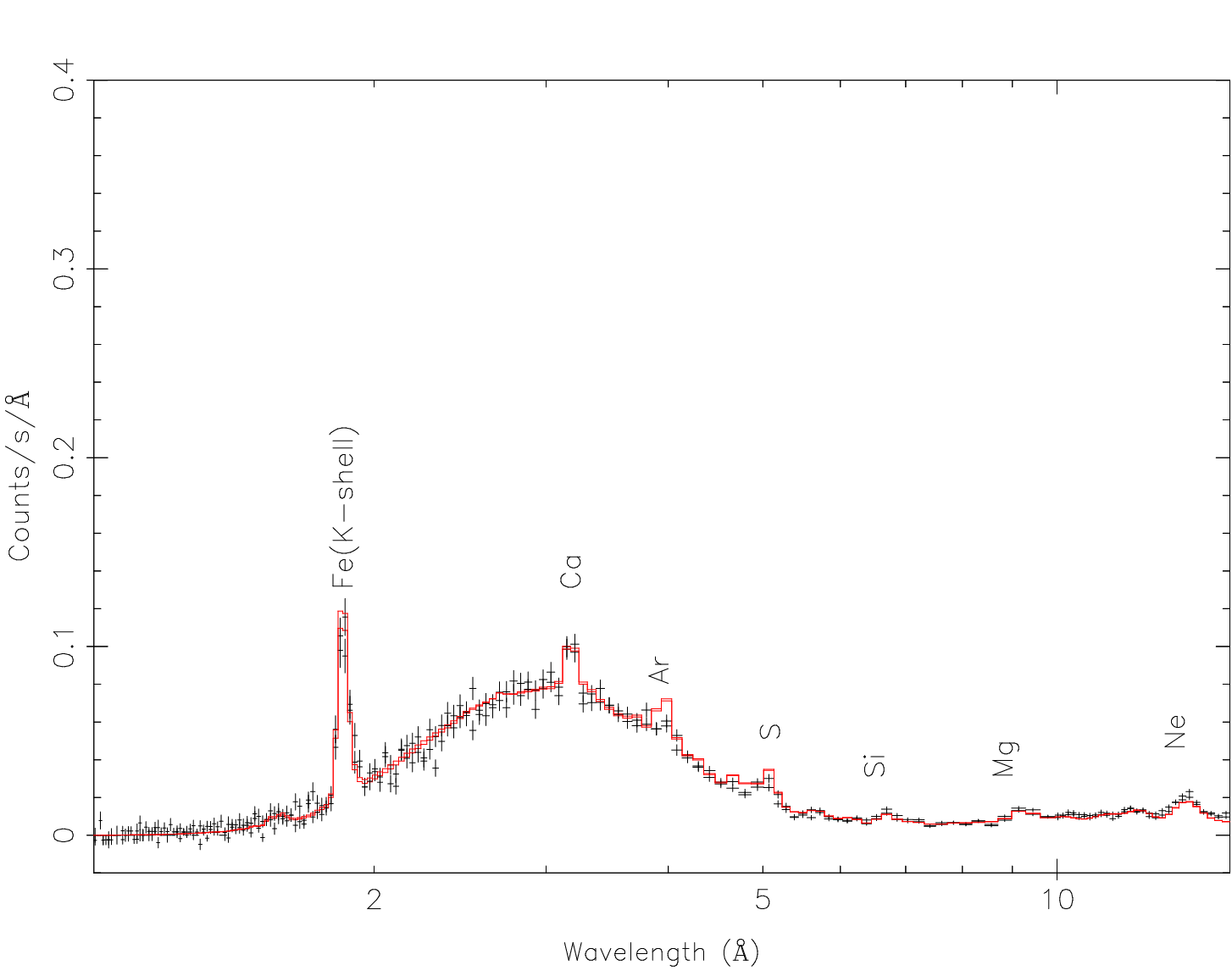,angle=-90,height=5cm,width=8.5cm}
}
\hbox{
\psfig{figure=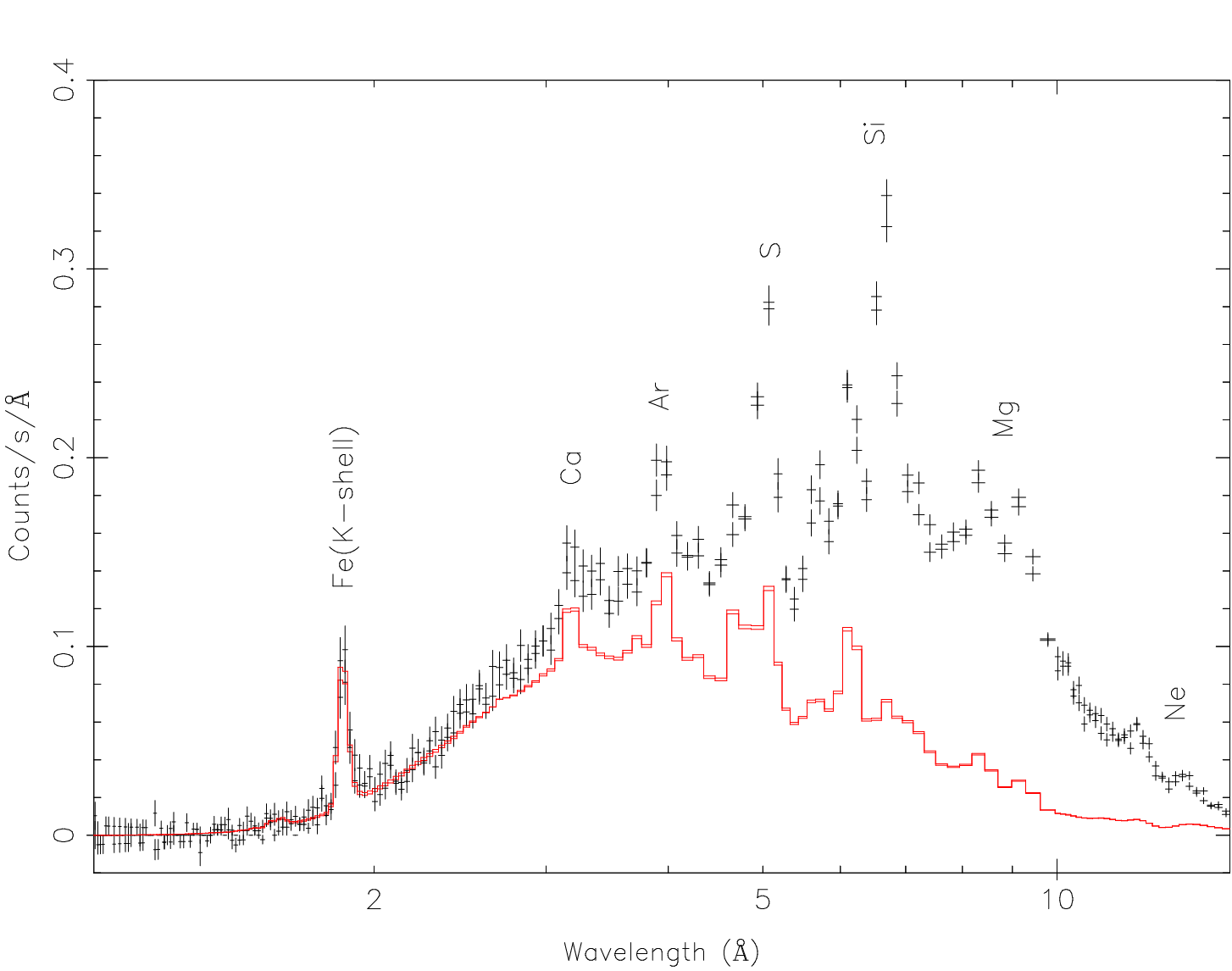,angle=-90,height=5cm,width=8.5cm}
\psfig{figure=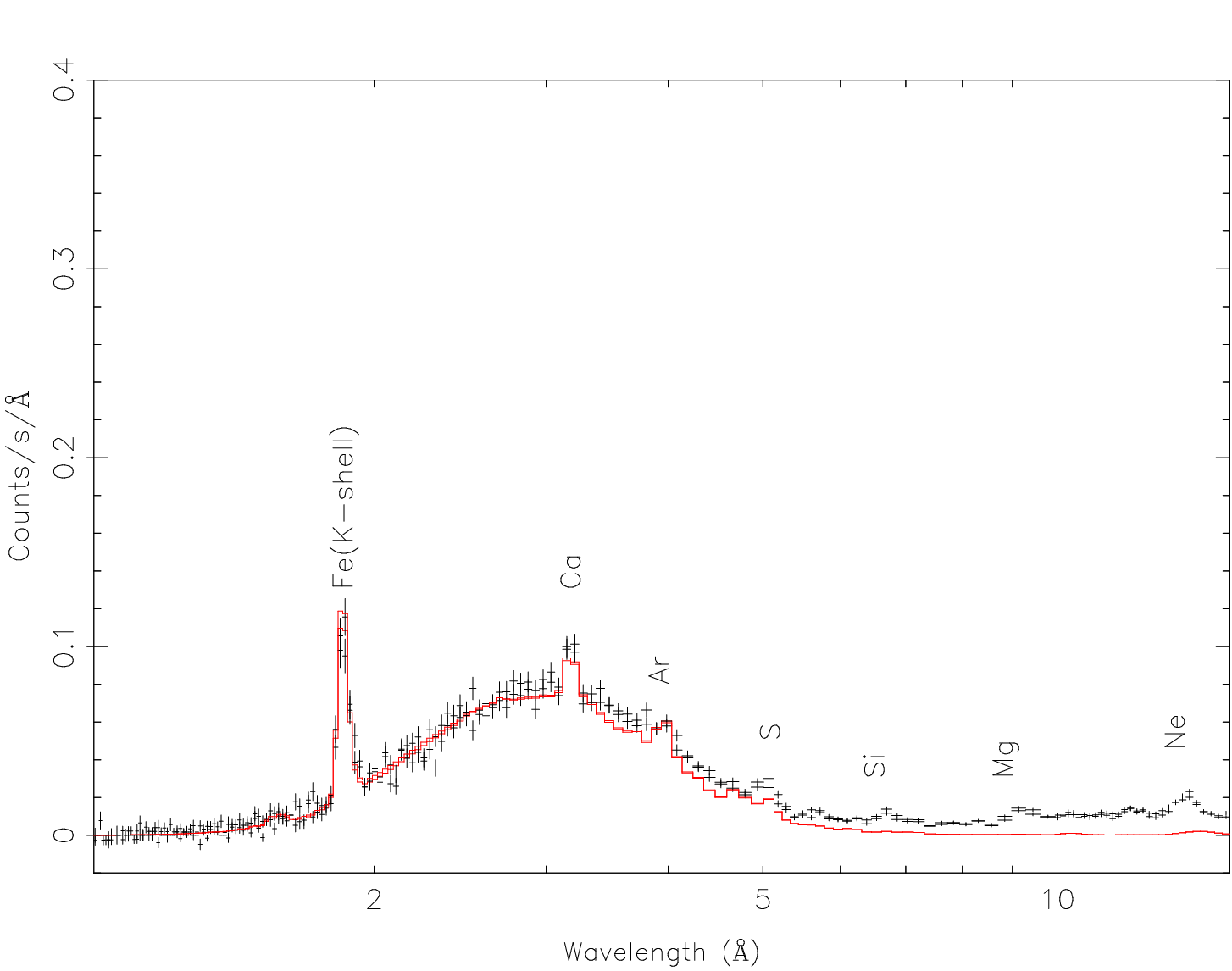,angle=-90,height=5cm,width=8.5cm}
}
\hbox{
\psfig{figure=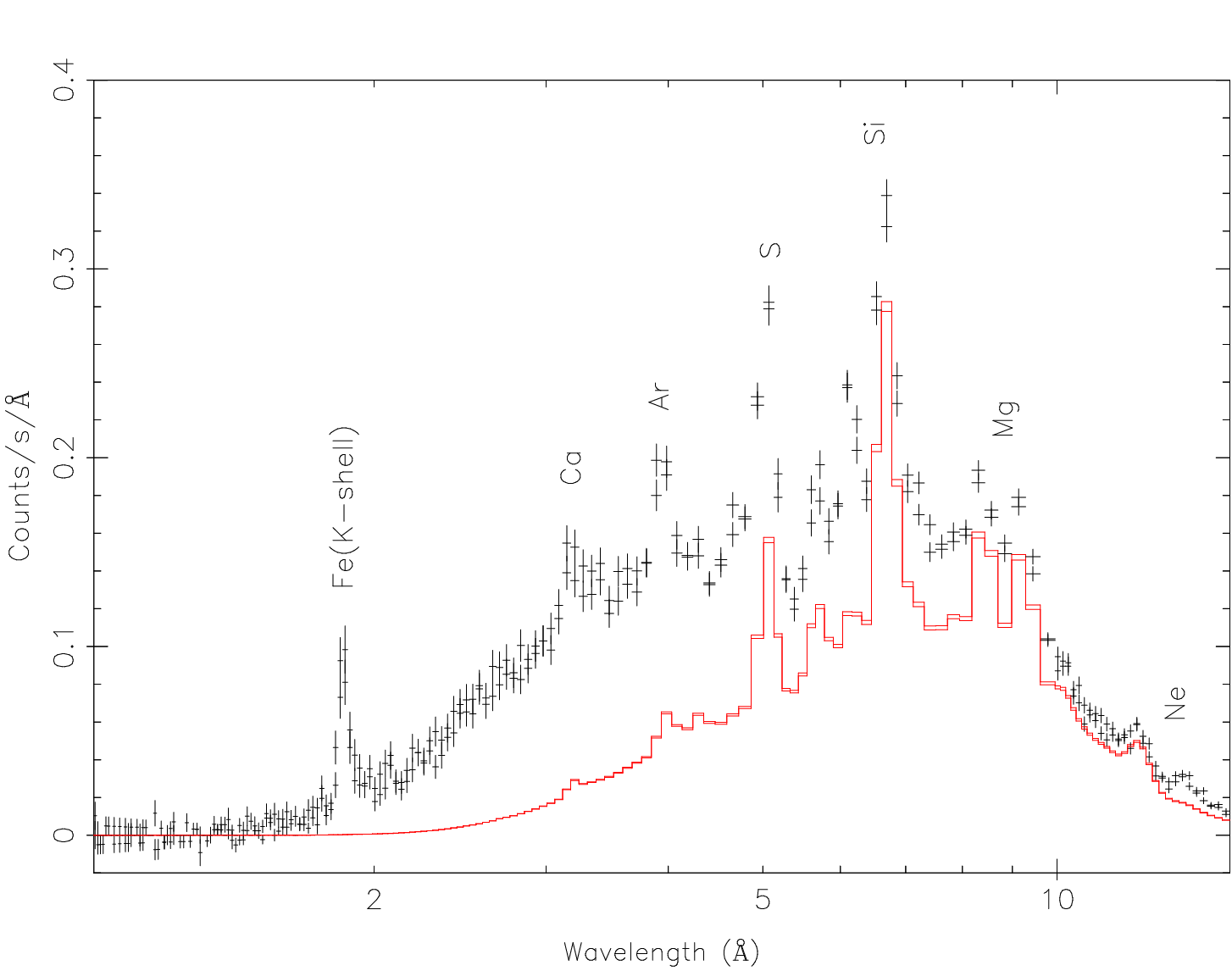,angle=-90,height=5cm,width=8.5cm}
\psfig{figure=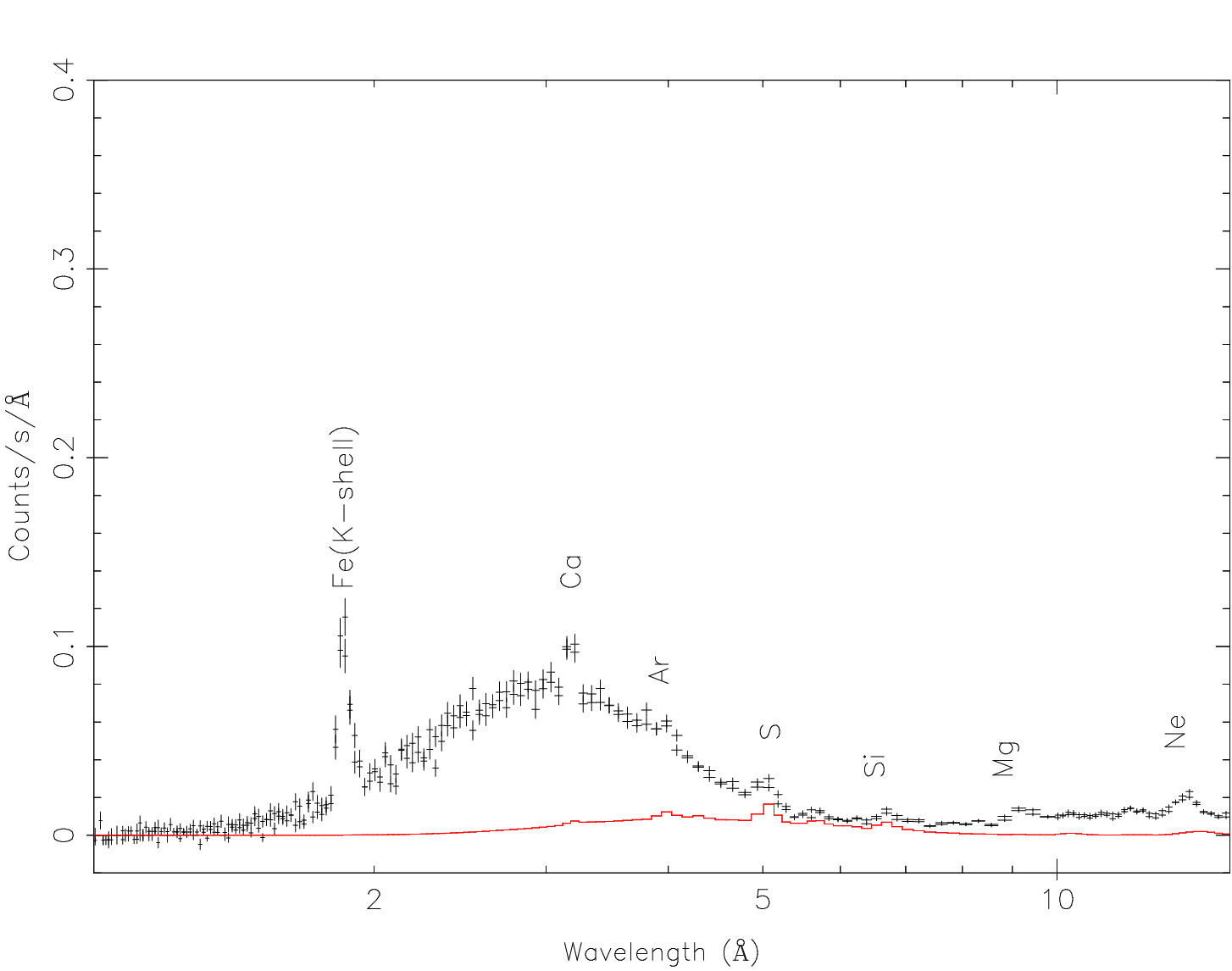,angle=-90,height=5cm,width=8.5cm}
}
\hbox{
\psfig{figure=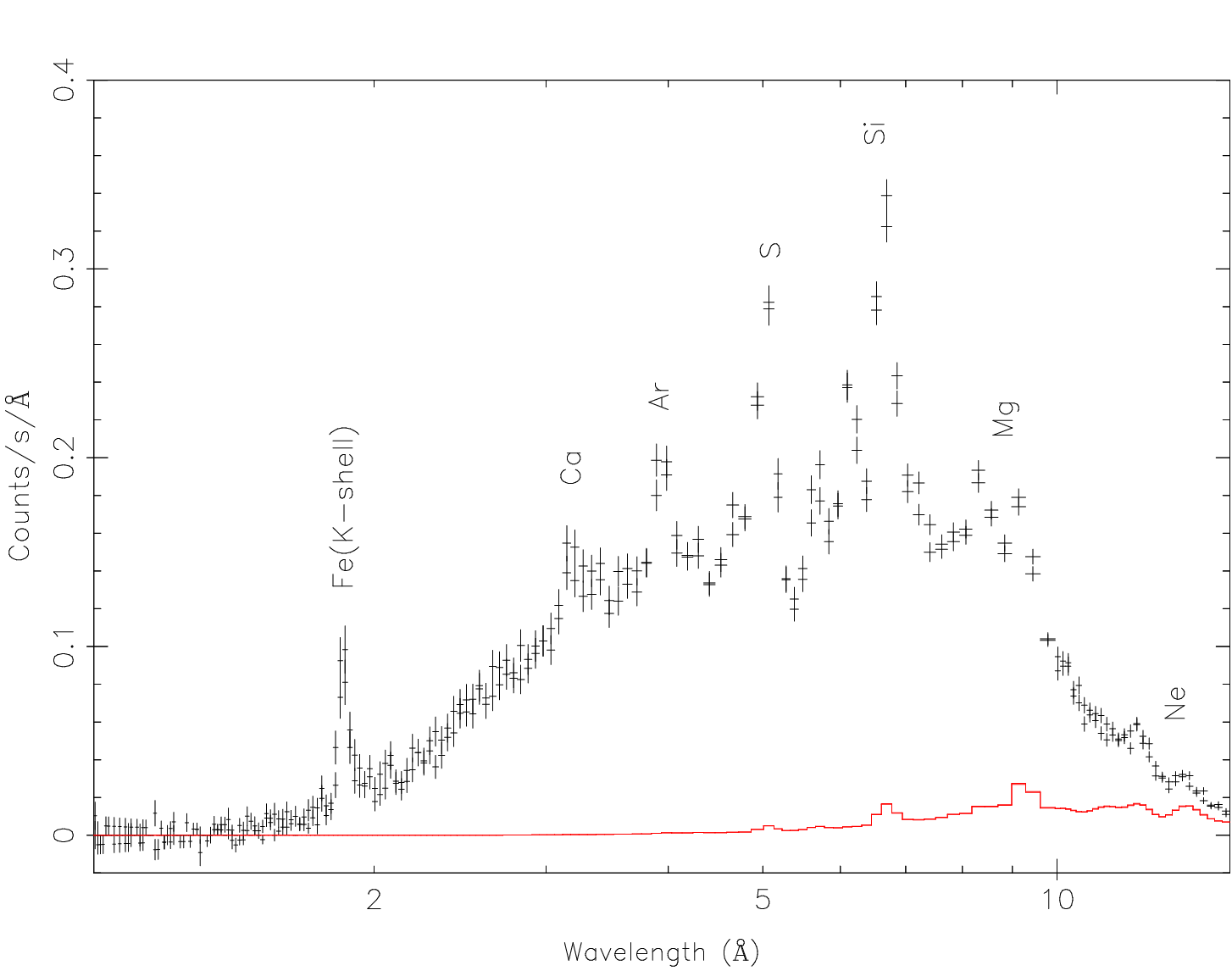,angle=-90,height=5cm,width=8.5cm}
\psfig{figure=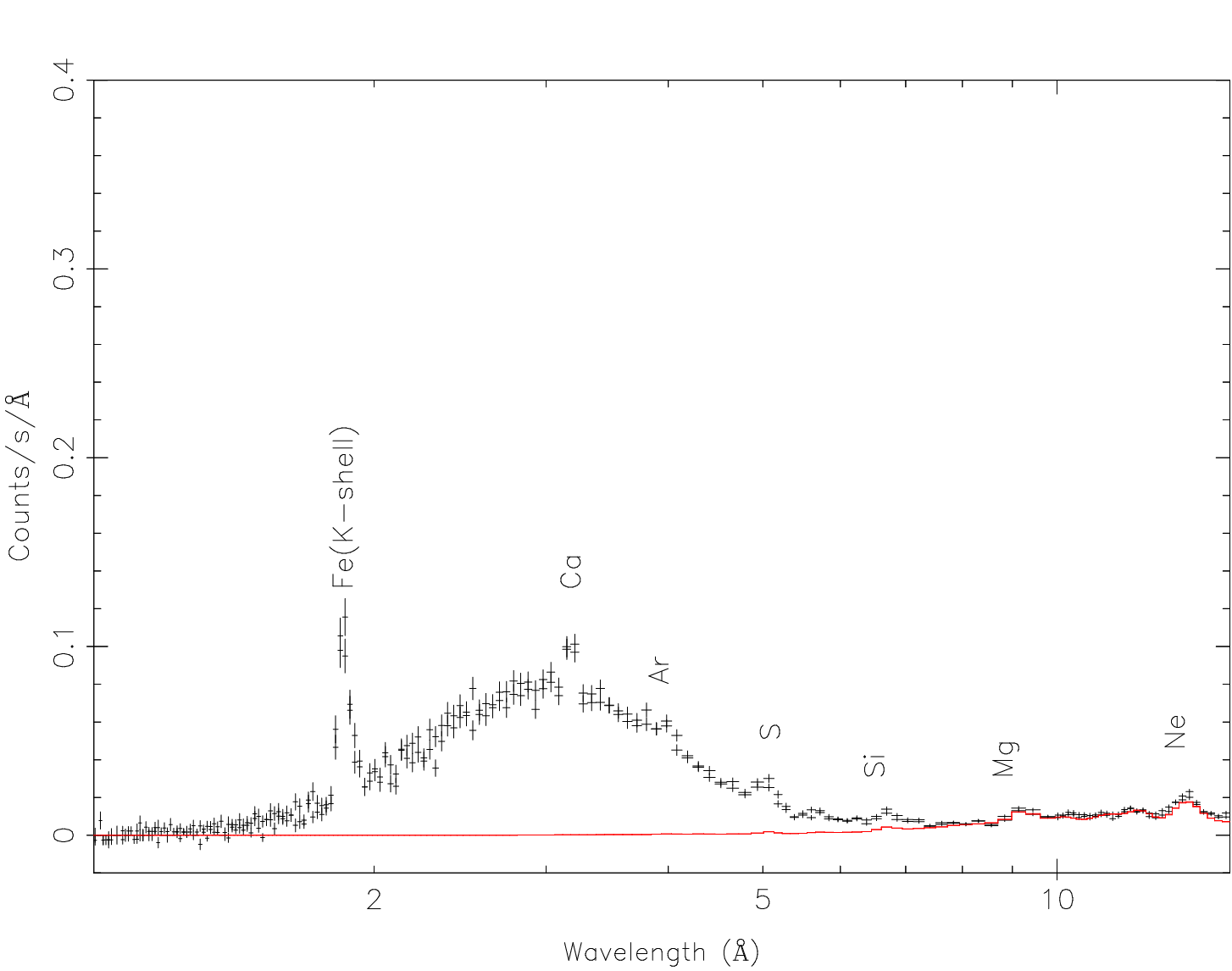,angle=-90,height=5cm,width=8.5cm}
}
\caption{Background-subtracted {\sl XMM}-{\sc epic-mos} spectra of
$\gamma^2$\,Vel at phases $\phi$\,=\,0.11 (high state; {\it left panels})
and $\phi$\,=\,0.37 (low state; {\it right panels}).  Error bars correspond
to 1$\sigma$ statistical errors including the effect of background.
The curves show from top to bottom our best-fit total model, and
the components 1, 2, and 3, respectively.  Some prominent lines are
labeled with the emitting ions.
        }
\label{compcomp.fig}
}
\end{figure*}

In Fig.\,\ref{fitspec.fig} we compare the combined MOS and the combined 
RGS spectra with the model while in Fig.\,\ref{compcomp.fig}, we show the various
contributions from the first three 
components of the best-fit
 model at both high and low state. 

%%%%% FIGURE 7 %%%%%

%%%%% TABLE 5 %%%%%
\begin{table}[t!]
\caption{Best-fit model line fluxes at Earth in units of
10$^{-13}$\,erg\,cm$^{-2}$\,s$^{-1}$ for the high state ({\it upper
section}) and low state ({\it lower section}) of $\gamma^2$\,Vel.  A
comparison with observations, and the relative contributions from the three
{\sc cie} components (1,2,3) and the photo-ionized component (4) are
also listed.
        }
\label{fitflux.tab}
\begin{center}
\begin{tabular}{r@{\ }|l@{\ }|c@{\ }|c@{\ }|c@{\ }c@{\ }c@{\ }c@{\ }r@{\ }  }
\hline\hline
                &         &       &                                     &     &     &     &         \\
$\lambda$ (\AA) & ion     & flux  & ${{{\rm obs.}^a}\over {\rm model}}$ & \multicolumn{5}{c}{contribution (\%)} \\
                &         &       &                                     &  1  & 2  &  3  &   4      \\
                &         &       &                                     &     &    &     &          \\[-1mm]
\hline\hline
                &         &       &                                     &     &    &     &          \\[-1mm]
 1.87           & \Fek    &  2.7  & 1.0                                 & 100 &  0 &   0 &   0      \\[+0.3mm]
 3.02           & \Caxx   &  0.05 & --                                  & 100 &  0 &   0 &   0      \\[+0.3mm]
 3.18           & \Caxix  &  0.86 & 0.93                                &  89 & 11 &   0 &   0      \\[+0.3mm]
 3.73           & \Arxviii&  0.23 & --                                  & 100 &  0 &   0 &   0      \\[+0.3mm]
 3.95           & \Arxvii &  1.3  & 1.31                                &  77 & 23 &   0 &   0      \\[+0.3mm]
 4.73           & \Sxvi   &  1.6  & 0.94                                &  97 &  3 &   0 &   0      \\[+0.3mm]
 5.05           & \Sxv    &  5.9  & 1.08                                &  41 & 59 &   0 &   0      \\[+0.3mm]
 5.68           & \Sixiii &  0.95 & 0.63                                &  11 & 89 &   0 &   0      \\[+0.3mm]
 6.18           & \Sixiv  &  2.   & 0.93                                &  61 & 39 &   0 &   0      \\[+0.3mm]
 6.68           & \Sixiii &  8.0  & 1.15                                &   8 & 92 &   0 &   0      \\[+0.3mm]
 8.42           & \Mgxii  &  3.1  & 0.65                                &   7 & 93 &   0 &   0      \\[+0.3mm]
 9.23           & \Mgxi   &  2.2  & 1.27                                &   1 & 95 &   4 &   0      \\[+0.3mm]
10.24           & \Nex    & 0.62  & 0.77                                &   2 & 95 &   3 &   0      \\[+0.3mm]
12.13           & \Nex    & 1.1   & 1.14                                &   0 & 91 &   9 &   0      \\[+0.3mm]
13.60           & \Neix   &  0.60 & 1.33                                &   0 & 12 &  75 &  13      \\[+0.3mm]
15.01           & \Fexvii & 0.23  & 0.96                                &   0 & 42 &  58 &   0      \\[+0.3mm]
33.74           & \Cvi    & 0.35  & 0.83                                &   0 &  0 &   0 & 100      \\[+0.3mm]
                &         &       &                                     &     &    &     &          \\[-1mm]
\hline
                &         &       &                                     &     &    &     &          \\[-1mm]
13.60           & \Neix   & 0.90  & 1.26                                &   0 &  0 &  92 &   8      \\[+0.3mm]
15.01           & \Fexvii & 0.29  & 0.69                                &   0 &  0 & 100 &   0      \\[+0.3mm]
18.97           & \Oviii  & 0.38  & 0.47                                &   0 &  0 &  72 &  28      \\[+0.3mm]
21.80           & \Ovii   & 0.22  & 1.10                                &   0 &  0 &  11 &  89      \\[+0.3mm]
33.74           & \Cvi    &  0.21 & 1.33                                &   0 &  0 &  0  & 100      \\[-0.7mm]
                &         &       &                                     &     &    &     &          \\
\hline\hline
\end{tabular}
\end{center}
\begin{flushleft}
{
\begin{description}
\item $a$: Observed fluxes from Tab.\,\ref{lineflux}.
\end{description}
}
\end{flushleft}
\end{table}

The resulting fit is quite satisfactory and confirms our previous
interpretation that the low state spectrum is an absorbed version of the
high state spectrum.  Components 1 and 2 turn out to have the same
temperature at high and low state but have a strongly increased absorption
column at low
state. As the interstellar column density is only
0.008$\times$10$^{22}$\,cm$^{-2}$ (Stevens et al. 1996)
most of the absorption must be due to circumstellar
material. We conclude that the hot
components 1 and 2 are deeply embedded in the WR wind at low state.

The soft and rather weak component 3 also suffers absorption but
much less than components 1 and 2. This cool CIE component must lie further
away from the WR star than components  1 and 2.

The observed spectrum in the wavelength region $\ga$\,20\,\AA\  ($<$ 1\,keV) is largely
due to a cool, recombining plasma (component 4) which is unaffected by
absorption. The low absorption indicates that it is emitted largely
outside of any dense material of the WR wind.

\subsubsection{Observed and predicted line fluxes}

Using the best-fit parameters (cf. Tab.~\ref{fitres.tab}) and the
absorption data by Morrison \& McCammon (1983), we have calculated the
predicted fluxes at Earth for a number of prominent lines (mainly of H- and
He-like ions), together with the relative contributions from the three {\sc
cie} components and the recombination component (Tab.~\ref{fitflux.tab}).
The agreement between observed and calculated fluxes is quite satisfactory,
except for the \Oviii line, which is a factor two larger than the observed
one.  The difference for this line could be reduced by a lower oxygen
abundance in component 3.

We note that component 1 is dominant in the formation of the lines
shortward of 5\,\AA, component 2 for the lines between 5 and
$\sim$\,12\,\AA, whereas components 3 and 4 emit the lines at longer
wavelengths. The Neon lines originate in components 2 and 3.

%%%%% FIGURE 8 %%%%%
\begin{figure}[t!]
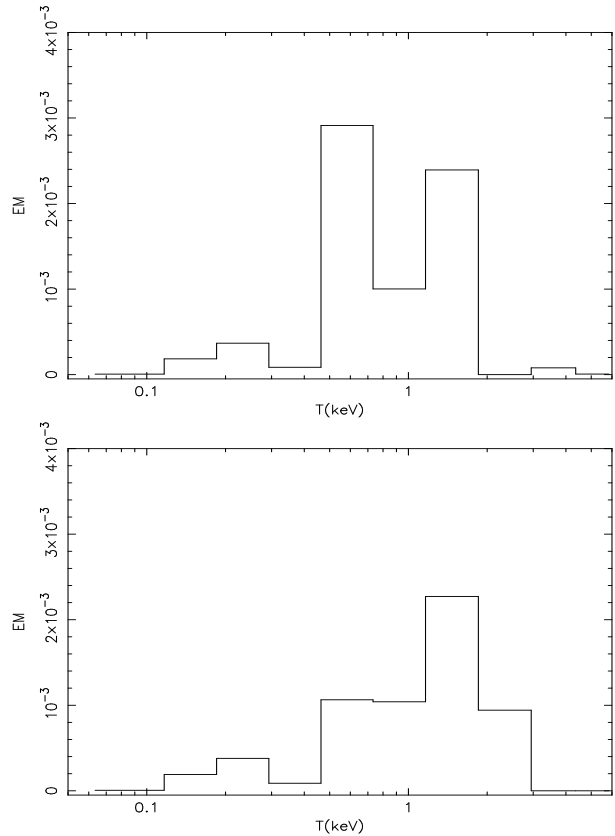

\hbox{\psfig{figure=35fig8to.ps,angle=-90,height=5.5cm,width=8.0cm,clip=}}
\hbox{\psfig{figure=35fig8bo.ps,angle=-90,height=5.5cm,width=8.cm,clip=}}
\caption{DEM modeling of the {\sl XMM}-{\sc epic-mos} and -{\sc rgs}
spectra of $\gamma^2$\,Vel ({\it top:} high state; {\it bottom:}
low state).  The emission measure EM per logarithmic temperature bin
is in units of 10$^{58}$\,cm$^{-3}$.
        }
\label{dem.fig}
\end{figure}

\subsection{DEM modeling}

Apart from multi-$T$ fitting, the \xray spectrum was also fitted with a {\sc
DEM}-modeling procedure. We define the differential emission measure
$DEM$ by $n_en_{\rm H}dV/d{\rm
  log}T$ (integrated over one logarithmic temperature bin $EM$ =
$n_en_{\rm H}V$).
The method is based on a spectral description by means of
a spline-method, and offers the possibility to obtain the differential
emission measure distribution and the abundances, simultaneously and for
different $N_{\rm H}$ values. The resulting continuous DEM
distribution is shown in  Fig\,\ref{dem.fig}. The DEM results
are consistent with those of the previous 3-T fitting
and therefore, we do not give them explicitly in a table.
The two narrow peaks in the EM distribution at high state are probably
artefacts due to
finite signal/noise and inaccuracies in the atomic data.

\section{Discussion}

\subsection{Wolf-Rayet mass loss and wind clumping}

The fact that the winds of Wolf-Rayet stars are clumped is now generally
accepted  and supported by a variety of
observations (e.g., Hillier 2003). Single WR stars
show random variability in their broad band flux, polarization and
also in emission line profiles (e.g., Moffat et al. 1988, Robert et
al. 1991, L\'epine \& Moffat 1999, Rodrigues \& Magalhaes 2000;
L\'epine et al. 2000, Marchenko 2003, and references therein).
These variations are most easily
interpreted in terms of density enhancements in the wind.
In  the case of $\gamma^2$\,Vel, L\'epine et al. (1999) reported
emission line profile variations that can be attributed to WR wind
inhomogeinities. Eversberg et al. (1999) concluded from their
variability observations that the whole WR wind 
is affected by clumping
and that the inhomogeinities propagate outward.  

In classical WR atmosphere models, the effects of
wind clumping and the mass loss
rate are difficult to disentangle. Different combinations 
of the two can lead to equally good descriptions
of the emission line profiles. It is thus useful to try and
obtain independent information about e.g. the WR mass loss rates
and then use this in conjunction with WR atmosphere models
to pin down clumping more accurately.

Here we use the \xray absorption to measure directly an absorption
column which yields a mass loss rate of 8$\times$10$^{-6}$\Msolar/yr
for the WC8 star in the system. This value is based on simple physics
and does not depend on any clumping model. It depends
linearly on the density. Our mass loss rate is in very good
agreement with a value from  polarimetric observations which yields 
7$\times$10$^{-6}$\Msolar/yr (Schmutz et al. 1997, St.-Louis et
al. 1988).  This rate is based on a scattering model and is also
proportional to the density. Other mass loss rates based
on methods that are proportional to the density squared (line
profile fitting, radio data) are higher, typically by a factor of 
four. Such methods are much more sensitive to clumping and if
clumping is included, the mass loss rates become smaller. With
our low mass loss rate from the \xray absorption we add another 
independent piece of evidence for lower WR mass loss rates.

Although the observational evidence for clumping is clear, it
is difficult to convert it into a physical model.  A simple description in
terms of a clumping factor $f$ was introduced by Schmutz (1997)
and Hamann \& Koesterke (1998). The clumping factor is defined as the
ratio of the density in a clump to the density in a homogeneous
 model with the same mass loss rate. Line profile fitting that
includes electron line wings provide values of typically $f\sim$ 10
with, however, a large uncertainty such that values in the
range 4\,-\,25 are possible (e.g., Morris et al. 2000). Our new mass loss
rate in conjunction with a WR atmosphere model is much more sensitive 
and provides a clumping 
factor $f\approx$\,16 for the WC8 star in $\gamma^2$\,Vel. 
If the interclump medium is assumed to be empty the volume filling
factor is $1/f \approx$ 0.06.

\subsection{Modeling the X-ray emission}

Our \xray spectra in both, low and high state, can be described by three relatively
hot {\sc cie} components with a temperature range from about 0.25 to
1.6 keV, and a cold photo-ionized component with a WC star composition.

Comparing our spectral fitting results with those obtained for the high
state by Skinner et al. (2001) with {\sl Chandra}-{\sc hetgs} and by Rauw
et al. (2000) with {\sl ASCA}, we find good agreement for temperatures,
emission measures and column densities (see Tab.~\ref{fitres.tab}).  For the
intrinsic X-ray luminosity Skinner et al.  (2001) obtained
7.8$\times$10$^{32}$\,erg\,s$^{-1}$ in the 0.4\,-\,10\,keV band at the
binary phase $\phi$\,=\,0.08, while Rauw et al.  (2000) obtained
9.0$\times$10$^{32}$\,erg\,s$^{-1}$ in the 0.5\,-\,10\,keV band. This
is in reasonable agreement with our value of
1.5$\times$10$^{33}$\,erg\,s$^{-1}$ at
high state.  At low
state, however, our fitting results are different from those obtained by
Rauw et al. (2000), but this may be explained by the fact that fits of
\ASCA lines cannot accurately constrain the emission measure and column
density of the second component.

\subsubsection{Intrinsic \xray emission from the binary stars?}

We have compared our $\gamma^2$\,Vel results with a 2-component model
of the O9.7Ib star $\zeta$\,Ori that was fitted to {\sl XMM}-{\sc rgs} data
(Raassen et al. 2004).  
From the relatively low emission measures of $\zeta$\,Ori
we expect that the `single-star' contribution from the O-type component in
$\gamma^2$\,Vel cannot play a significant role in the overall X-ray
emission.  Some contribution to
the soft and relatively weak component 3 due to shocks in the wind of
the O star can, however, not be excluded.

From the $L_{\rm x}$-$L_{\rm bol}$ relation for single
O-type stars derived from \ROSAT data (Bergh\"ofer et al.  1997),
we estimate a contribution to the intrinsic X-ray luminosity (in the
0.1\,-\,2.4\,keV band) of about 1.2$\times$10$^{32}$\,erg\,s$^{-1}$ with an
error of a factor of $\sim$\,2\,-\,3.  In principle, this agrees with our
fitting results for component 3.

Oskinova et al. (2003) gave for single WC stars upper limits to the
X-ray luminosities of 0.025\,-\,0.32 10$^{32}$\,erg\,s$^{-1}$ in the 
0.2\,-\,10\,keV band. They explained the apparent absence of X-rays from single WC
stars by strong absorption in the stellar wind. Accordingly stellar
wind shocks from the WC star
appear not to contribute significantly to the observed X-ray
luminosity in $\gamma^2$\,Vel.

\subsection{Towards constraints for hydrodynamical models}

For makers of hydrodynamical wind collision models, $\gamma^2$\,Vel is a
interesting object because its easy observational access can
potentially provide
important constraints. The broadband \xray light curve as observed with
\ROSAT (Willis et al. 1995) has already yielded information about the
orientation and opening
angle of the wind cavity behind the O star. A similar light-curve
obtained 
with high spectral resolution will reveal the detailed geometric structure
of the wind-wind collision zone. Already our observations taken at only two
phases show that a low temperature component exists that must be much more
extended than expected. A tomographic survey, covering a complete
orbit will reveal the distribution of the \xray luminous matter as well as
the absorption column in various directions.

An observational quantity that is of particular interest to model makers
is the elemental abundances.  Here we report for the first time a
neon abundance that is enhanced compared to solar. This demonstrates that
the shocked material is at least partly from the WC star. Theoretically,
it is expected that the WC material dominates the  \xray emission.

In this context it may be useful to re-evaluate the cooling parameters
 that characterize the  wind-wind collision zones because
 orbital parameters and mass loss rates have recently been revised.
 The cooling parameter $\chi$ (Stevens,
 Blondin \& Pollock 1992) is given by the ratio of the cooling and the
 dynamical time scale
 \be
 \chi = \frac{t_{\rm cool}}{t_{\rm dyn}}= \frac{D_{12}\,v_{8}^4}{\Mdot_{-7}}
 \ee
 \noindent
 where $\Mdot_{-7}$ is the mass loss rate in [10$^{-7}$\,M$_\odot$yr$^{-1}$],
 $D_{12}$ is the distance from the star center to the contact discontinuity
 in [$10^{12}\,\rm{cm}$], and $v_{8}$ is the wind velocity in
 [$10^8\,\rm{cm\,s^{-1}}$].  With the new lower mass loss rates for
 the clumped 
WR wind (which is similar to the one derived in this paper)
and the O-type component 
from De Marco et al. (2000),
 the value of $\chi$ turns out to be $\sim$\,1 for the shocked WC material and
 $\sim$\,40 for the shocked O star material.  In terms of this parameter, the
 shocked O-type star wind can be considered adiabatic, while for the shocked
 WC wind
 radiative cooling begins to be important.

For an adiabatic wind-wind
collision, the ratio of the X-ray luminosities emitted by the shocked winds
is

\be
\frac{L_{\rm x}^{\rm WR}}{L_{\rm x}^{\rm O}}\approx
(\frac{v_{\rm O}}{v_{\rm WR}})^{5/2}
\ee

With the observed terminal wind velocities of the O  and the WC
star of $v_\infty$\,=\,2500 and 1450\,km\,s$^{-1}$, respectively, this
luminosity ratio is 3.3.  The X-ray emission is expected to be
dominated by shocked WC material and should reflect
the abundance pattern of the WC star. If the observations would show a
significant contribution from material with solar abundance (i.e. from the O
star), this could be caused by a slower O star wind because e.g.
the O star material can not reach the terminal velocity before it is
entering the collision zone (e.g. Pittard \& Stevens 1997 and references
therein).

\subsection{On the interpretation of the \Cvi and \Cv recombination features}

The existence of narrow radiative recombination continuum features from
hydrogen-like and helium-like carbon in the spectrum of $\gamma^2$\,Vel implies that
highly ionized carbon is in the presence of cold electrons, and is recombining
with them.

The most obvious explanation for this is that some part of the WR wind
is being photoionized by the hard X-ray emission from the colliding wind
shock.

The ionization parameter expected in the WR wind can be
computed by approximating the X-ray emission as a point source at the
wind collision point colinear with the two stars. If the wind is smooth,
\[ n_{e}=\frac{\dot{M}}{4\pi\mu m_{p}r^{2}v} \] 
The ionization parameter is
\[ \xi=\frac{L_{x}}{n_{e}r_{x}^{2}}=\frac{4\pi\mu
m_{p}v_{\infty}L_{x}}{\dot{M}}(\frac{r}{r_x})^{2}w(r)=\xi_{mid}(\frac{r}{r_x})^{2}w(r)\] 
where $r_{x}$ is the distance from the colliding wind shock apex. We also make
use of the notation $v(r)=v_{\infty}w(r)$, where
$w(r)=(1-\frac{r}{R_{WR}})^{\beta}$ is almost one at radii more than a few
$R_{WR}$. The midplane is defined to be the plane which is equidistant from 
the WR star and the shock. At midplane, 
\[ \xi_{mid}=\frac{4\pi\mu m_{p}v_{\infty}L_{x}}{\dot{M}}=1.2\times
10^{-2}\, \rm{erg\, cm^{-1}\, s^{-1}} \] 
or log $\xi_{mid} = -1.9$ (taking $L_{x}=10^{33}\, \rm{erg\, s^{-1}}$,
$\mu=2$, $\dot{M}=8\times 10^{-6}\, \rm{M_{\odot}\, yr^{-1}}$, and
$v_{\infty}=1500\, \rm{km\, s^{-1}}$). 

The ionization parameter in the WR wind should be within a factor of 3 of
$\xi_{mid}$ throughout most of the volume. Close to the WR star it will be
lower due to the higher densities, but we are not able to see that region
due to photoelectric absorption. Close to the shock, the ionization parameter
will be higher, but this is a relatively small amount of material.
For a more detailed treatment of the geometrical dependence of the ionization
parameter, see \cite{hmc77}, \cite{lp96}, and \cite{s99}. These deal with 
the photoionized winds of high mass X-ray binaries, but the mathematical
description is similar. 

The value of $\xi$ we calculate at the midplane for a smooth wind is three
orders of magnitude smaller than the value required to completely strip carbon
and allow us to observe \Cvi emission ($\rm{log}\,\xi = 1$). In addition to
this, we know that the \Cvi emission does not appear to be phase variable, at
least at the two phases we observed. Taken together, this implies that if the
recombination emission comes from the photoionized WR wind, it comes from far
out, and in a very rarefied part. If this is the case, then the \Cv and \Cvi
emission is coming from the interclump medium, which is highly rarefied
compared to the density expected for a smooth wind. This is certainly
plausible considering the degree of clumping which is known to be present in
WR winds in general, and the clumping we infer in the wind of $\gamma^2$\,Vel.

The other possible explanation is that the hot plasma created in the colliding
wind shock stops cooling radiatively at some point before carbon recombines
(presumably the density is falling off pretty fast, and adiabatic cooling
should become more important at some point), and the highly ionized carbon
ions are either mixed with cool electrons in the WR wind further out in the
flow, or the adiabatic cooling allows the electrons in the colliding wind flow
to become cool enough to reproduce the observed effect when they do eventually
recombine. 

One other important piece of information comes from the rest of the
spectrum. We see emission from Ne\,{\sc ix}, Fe\,{\sc xvii}, O\,{\sc viii}
and \Ovii which we
believe comes from material far out in the post-collision flow, mainly because
these lines have the same strength at both phases observed, but also because
the absorbing column must be very low for us to see these lines at all. We
would expect to see some kind of emission from \Cvi and \Cv from this same
material, because it must cool to the ambient temperature of the wind
eventually, unless the emission occurs at a radius where the WR wind is still
optically thick at about $30\,{\rm{\AA}}$. Even if mechanisms other than
radiative cooling are important, the fully stripped carbon atoms created in
the colliding wind shock must recombine eventually, and when they do they must
emit one X-ray photon per atom per electron added. However, the continuum
optical depth (mainly from \Civ  in the cool WR wind) is higher for \Cvi
emission than for Ne\,{\sc ix}, so it is possible that \Cvi emission in the
post-shock flow would be absorbed. 

If we believe that the observed carbon emission is just from the cooler parts
of the post-collision flow, we should ask why we see different emission
mechanisms at work in the case of carbon as opposed to neon, iron and
oxygen. The observed line ratios are \Neix is consistent with what one would expect
in a hot, collisionally ionized plasma. The statistics for \Fexvii and \Fexviii
and \Ovii and \Oviii are relatively poor, and it is not impossible that \Ovii
and \Oviii RRCs are present in the spectrum and are merely blended with the
iron L-shell emission. Clearly at some point, the cooling of the post-shock
flow switches from primarily radiative to primarily adiabatic, but it is not
clear whether this can account for the recombining \Cvi and C\,{\sc v}.

\section{Summary and Conclusions}

High-resolution X-ray spectra obtained at different orbital phases provide
a wealth of information about $\gamma^2$\,Vel.  Modeling the X-ray emission
constrains the physical structure of the wind-wind
collision zone, whereas the
absorption observed at non-maximum phases gives indications about
the geometric distribution of the emitting as well as the
non-emitting
material.  Both, emission and absorption
are important and reveal different but linked aspects of the $\gamma^2$\,Vel
system. It is indeed likely that a comprehensive tomographic analysis
using \xray spectra taken at many more orbital phases will allow
a detailed mapping of the colliding wind region as well as of
the ambient material. In particular, the hypothesis of a constant clumping
factor around the orbit could be tested.

\subsection{WR mass loss and wind clumping}

Phase dependent \xray emission from  $\gamma^2$\,Vel can be used to
analyze the Wolf-Rayet wind. 
In order to quantitatively interpret the absorption at low state,
we apply a previously published WR model atmosphere with a
smooth density distribution (De\,Marco et al 2000). 
This model atmosphere is the result from a fit to the broad WR emission lines.
The column
density required by the observed \xray absorption is a factor of 4 lower than 
what is predicted by this model.
The mass loss rate that matches the \xray absorption is correspondingly
smaller. We conclude that the WC8 star in  $\gamma^2$\,Vel
loses mass at a rate of only 8$\times$10$^{-6}$\Msolar/yr.
 
The discrepancy between our directly measured mass loss rate and
the one required by the model atmosphere can be reconciled if the
wind is clumped.
In order to still fit the WR emission line spectrum
with the reduced mass loss a clumping factor f = 16 is required. 

%%%%% FIGURE 9 %%%%%
\begin{figure}
\hbox{\hskip-0.2cm\psfig{figure=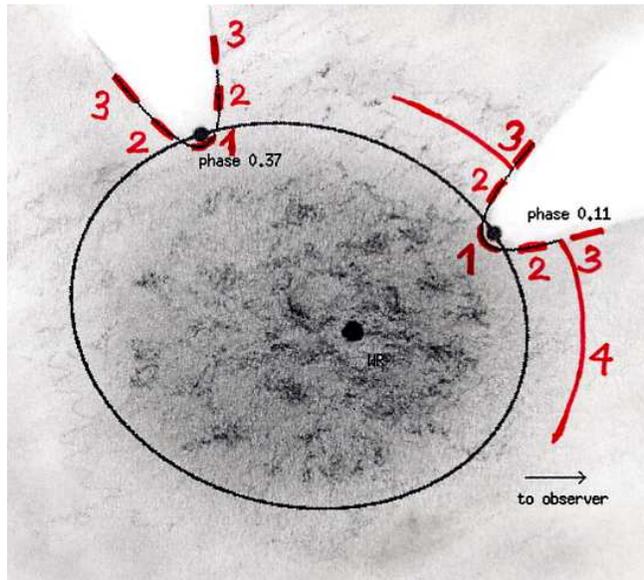,height=7.7cm,width=8.5cm,clip=}}
\caption{Sketch of the $\gamma^2$\,Vel system (same as Fig.\,1) with the
probable locations of our components 1 to 4. The Wolf-Rayet star
is at the center.}
\label{figure9}
\end{figure}

\subsection{Size of line emitting regions}

The observed absorption behaviour also constrains the geometry of
the \xray line emitting region. It is very interesting that in our
spectra the separation between absorption and no absorption is
quite sharp. While the \Neix lines remain unabsorbed the \Nex lines are
reduced by a factor of 5 (see Tab. 2). In terms of temperature this
means
that the plasma hotter than 5 MK is heavily absorbed at phase 0.37
while the cooler
plasma is not.  This is also reflected in our emission model in which the
components 1 and 2 with temperatures of 8 and 19 MK are strongly absorbed
whereas component 3 with a temperature of 3 MK is only weakly absorbed.
We conclude that components 1 and 2 are formed in 
the central part of the colliding winds which is deeply embedded in
the WR wind.
The cool (3\,MK) component is clearly detached from this hot region
(see Fig.\,9).

Furthermore, the
\Neix lines that predominantly come from this region are not affected by the
UV radiation of the O star. They either  are formed far away from the O
star or they are shielded from that UV radiation by intervening material.
In either case, the O star is not likely to contribute much to them
and we conclude that firstly, this component is associated with the colliding
winds and secondly that this region must be rather extended for it to still
be well detectable at phase 0.37.

\subsection{Neon abundance}
A further interesting piece of evidence comes from the neon abundance we
derive. Neon is
considerably enriched through nuclear processing in WC stars and therefore
differs significantly from solar abundance.  Neon is
in fact the only element that emits copious line radiation from the
collisionally ionized region, and that allows to discriminate between WC and solar
composition. We find a clear  Neon enhancement compared to solar which
indicates that Wolf-Rayet material is present at least in components 2
and 3.

It is noteworthy that much of what we learn about the collision zone actually
comes from the \Neix and \Nex lines. Apart from discriminating between WC and
solar abundance patterns they also provide a
dividing line between absorption and no absorption at phase 0.37. 
These lines seem to hold the key for further progress and their
behaviour at other phases should certainly be very interesting to follow.

\subsection{\xray emission variability}
An interesting feature of our \xray spectra is the high energy end.
The section above 4.5\,keV is little affected by intrinsic absorption and
interestingly there is
no observable difference between the two phases. In particular the
highest temperatures in the wind-wind collision zone seem to be the same
at both phases. This is remarkable because the binary separation
has changed from 0.83 AU at phase 0.12 to 1.27 AU at phase 0.37. From
a simple 1/D law one would expect a 50\% flux reduction (see e.g.,
Stevens et al. 1992). This confirms
the result of Rauw et al. (2000) that $\gamma^2$\,Vel  does not follow
a 1/D distance relation. The \xray  flux and temperature from the hottest
plasma as detected by \XMM is not affected by the orbital separation.

\subsection{Recombining plasma}
Apart from the shock excited components we also find a recombining plasma.
The relation (if any)  between this fourth component to the shocked material
is not clear. The recombining plasma is highly ionized and is not absorbed
at phase 0.37. It
therefore comes from far out in the binary system. We also know that it is
of very low temperature of about 40\,000\,K and that it is of WC composition.
Possibly this plasma is due to
photoionization through the X-rays from the wind-wind collision region. 
This radiation propagates through the rarefied and warped cavity behind the
O star and irradiates the higher regions of the WR wind. There it may
re-ionize some of the material.

\begin{acknowledgements}
We would like to thank R.~Walder and H.M.~Schmid for fruitful discussions.
The SRON National Institute for Space Research is supported financially by
NWO. M.A. and M.G. acknowledge support from the Swiss National Science Foundation
(fellowship 81EZ-67388 \& grant 2000-058827). M.A. and M.A.L. acknowledge support by
 a grant from NASA to Columbia University for XMM-Newton mission
 support and data analysis.

\end{acknowledgements}

\end{document}